\documentclass[
 reprint,
bibnotes,
amsmath,amssymb,amsfonts,
aps,
pre,
floatfix,
table
]{revtex4-2}

\usepackage{amsmath}
\usepackage{dsfont}
\usepackage{subfigure}
\usepackage{placeins}
\usepackage{graphicx}
\usepackage{dcolumn}
\usepackage{bm}
\usepackage{hyperref}
\usepackage[mathlines]{lineno}
\usepackage[normalem]{ulem} 

\usepackage{etoolbox}
\usepackage{xcolor}

\usepackage{tikz}  
\usetikzlibrary{svg.path}

\begin{document}


\title{Space and time correlations for diffusion models with prompt and delayed birth-and-death events}

\author{Th\'eophile Bonnet}
 \email{theophile.bonnet@cea.fr}
\author{Davide Mancusi}
 \email{davide.mancusi@cea.fr}
\author{Andrea Zoia}
 \email{andrea.zoia@cea.fr}
\affiliation{Universit\'e Paris-Saclay, CEA, Service d'Etudes des R\'eacteurs et de Math\'ematiques Appliqu\'ees, 91191, Gif-sur-Yvette, France}%

\date{\today}
\title{Space and time correlations for diffusion models with prompt and delayed birth-and-death events}

\begin{abstract}
Understanding the statistical properties of a collection of individuals subject
to random displacements and birth-and-death events is key to several
applications in physics and life sciences, encompassing the diagnostic of
nuclear reactors and the analysis of epidemic patterns. Previous investigations
of the critical regime, where births and deaths balance on average, have shown
that highly non-Poissonian fluctuations might occur in the population, leading
to spontaneous spatial clustering, and eventually to a ``critical
catastrophe'', where fluctuations can result in the extinction of the
population. A milder behaviour is observed when the population size is kept
constant: the fluctuations asymptotically level off and the critical
catastrophe is averted. In this paper, we shall extend these results by
considering the broader class of models with prompt and delayed birth-and-death
events, which mimic the presence of precursors in nuclear reactor physics or
incubation in epidemics. We shall consider models with and without population
control mechanisms. Analytical or semi-analytical results for the density, the
two-point correlation function and the mean-squared pair distance will be
derived and compared to Monte Carlo simulations, which will be used as a
reference.
\end{abstract}
\maketitle

\section{Introduction}

Many systems of interest in physics and life sciences can be
modeled in terms of a collection of individuals undergoing
random displacements coupled with birth-and-death events of
the Galton-Watson type \cite{harrisTheoryBranchingProcesses,
berg1993random, williamsRandomProcessesNuclear1974,
bharucha, bartlett1960stochastic,
pazsitNeutronFluctuationsTreatise2007}.
Two prominent examples concern for
instance the spread of epidemics and the propagation of
neutron-induced fission chains in nuclear reactors. For
epidemics, the displacements are those of the infected,
i.e., the carriers of the pathogen; births are due to the
appearance of new infected individuals due to contacts of
the infected with susceptibles; and deaths are due to the
transitions of the infected to the recovered (or possibly
back to the susceptibles) \cite{berg1993random}. For neutrons, displacements
result from free flights and from scattering collisions with
the nuclei of the
traversed media, leading to random re-orientations of the
flight directions; births are due to fission events, which
split the nuclei and emit extra neutrons;
and deaths are due to collisions leading to sterile
captures, whereupon the neutron life is terminated \cite{williamsRandomProcessesNuclear1974}.
In some models, a fraction of the birth events may occur
with a
delay: for epidemics, delayed infections are related to the
incubation time of the pathogen \cite{bailey}, whereas in reactor physics
the appearance of delayed neutrons is due to the
de-excitation of the fission fragments, which may emit extra
neutrons after exponentially distributed decay times \cite{williamsRandomProcessesNuclear1974}.

By virtue of their common ingredients, the mathematical
descriptions of epidemics and fission chains strikingly
share many features \cite{harrisTheoryBranchingProcesses},
although model-specific
traits must be taken into account in
realistic applications. The evolution of populations subject
to diffusion, births and deaths can be described by a
variety of frameworks, encompassing deterministic and
stochastic approaches, each having distinct advantages and
shortcomings \cite{kendall, bell_nuclear_1970}. When births are compensated by deaths, the
number of individuals is constant on average, and the system
is said to be critical \cite{harrisTheoryBranchingProcesses}: this regime is usually realized in
nuclear reactors, where physical parameters are precisely
tuned so that the energy released by fission is stationary \cite{williamsRandomProcessesNuclear1974}.
The critical regime also plays a central role in epidemics,
non-trivially affecting
control policies in the transition between the
super-critical and the sub-critical phases \cite{nature_epidemics}. If the number of
individuals is large, as in later stages of epidemics or in
nuclear reactor operated at full power, the
average behaviour of the population is sufficient to
completely characterize the system. On the
contrary, if the number of individuals is small, which is
the case e.g.~for epidemics during the early outbreak phases
and for nuclear reactors operated at low power, the
fluctuations around the average might be relevant and a
stochastic description is required \cite{williamsRandomProcessesNuclear1974,
bartlett1960stochastic}. In this case,
one either uses deterministic methods to solve the equations
for the statistical moments of the population, or Monte
Carlo methods to sample the random dynamics.

Although mean-field zero-dimensional approaches can provide
a condensed representation of the system dynamics,
stochastic models accounting for the full description of the
spatial behavior of the population are of utmost importance,
in that they are key e.g.~to characterizing the extent of
epidemic outbursts \cite{pnas_epidemics, spatial_epidemics, spatial_epidemics2}
or to detecting possible malfunctioning
or power tilts in nuclear reactors \cite{williamsRandomProcessesNuclear1974, 
pazsitNeutronFluctuationsTreatise2007, clusteringExp2021}. Modern state-of-the-art
Monte Carlo simulation codes allow addressing real-world
applications with unprecedented accuracy and have been
successfully used to compute the lower-order moments of the
populations, namely, the average  number of individuals
$\mathbb{E}[n_V](t)$ in a given spatial region $V$ at time
$t$ and the two-point correlation function
$\mathbb{E}[n_{V_1} n_{V_2}](t_1,t_2)$ between two regions
$V_1$ and $V_2$ at times $t_1$ and $t_2$
\cite{clusteringExp2021, tripoli_analog, mc_epidemics, mc_epidemics2}. The analysis
carried out with highly sophisticated simulation codes can
be effectively complemented by the investigation of
simplified mathematical models that avoid unnecessary
complication and
yet retain the key physical ingredients, possibly leading to
semi-quantitative predictions. In the
context of reactor physics, explicit expressions for the average
number of individuals and the two-point correlation function
of the population have been recently derived, for one such model
combining a Galton-Watson process with Brownian diffusion in
bounded domains \cite{zoiaClusteringBranchingBrownian2014,
houchmandzadehNeutronFluctuationsImportance2015,
zoiaNeutronClusteringSpatial2017}. Inspection of the
resulting solutions of the moment equations shows that, in
the critical regime, the birth-and-death events will induce
strongly non-Poissonian fluctuations in the number of
individuals being present at each spatial site, which lead
to a wild spatial patchiness (clustering) instead of the
expected uniform density. Eventually, the population will
face a ``critical catastrophe'', where fluctuations
can result in an extinction event. Furthermore, it has been
shown that the interplay between prompt and delayed
birth-and-death events affects the time scales of
clustering and of the critical catastrophe
\cite{houchmandzadehNeutronFluctuationsImportance2015,
zoiaNeutronClusteringSpatial2017}. Although these
investigations were mainly motivated by applications
pertaining to nuclear reactor physics, the main findings
would carry over to the spatial behaviour of epidemics, with
minimal modifications. For instance, it is known that around 
criticality all realistic epidemiological models
display large fluctuations that dominate the
dynamical behaviour of the system, regardless of the specific 
details of the infection process \cite{nature_epidemics}.

In the basic formulation of these models, the size of the population is free
to evolve according to the stochastic rules of the process, without any
constraint. In practice, epidemics are subject to health
policies and restrictions, and nuclear reactors are
similarly endowed with intrinsic
(physics-based) and external
control mechanisms to prevent power drifts. A simple, yet
effective way of mimicking some form of feedback acting
against variations in the number of individuals consists in
enforcing the population size to be exactly constant.
Such control mechanisms have been shown to have a dramatic
impact on the evolution of the fluctuations, compared to
the free case. In constrained models,
clustering is still present at the local scale, but
the spatial fluctuations at the global scale asymptotically
level off and the critical catastrophe is averted
\cite{mulatierCriticalCatastropheRevisited2015,
zoiaNeutronClusteringSpatial2017}. For the sake of
simplicity, so far such models with population control have only
addressed prompt birth-and-death events. Within the
framework of a simplified nuclear reactor model, in this 
paper we will extend these findings by introducing a
generalized approach to the enforcement of population control, where the
hypothesis of constant population size is relaxed and
delayed birth events are explicitly considered. Analytical
or
semi-analytical formulas for the neutron spatial density,
the two-point correlation function and the mean-squared pair
distance between particles will be derived and compared to
Monte Carlo simulations, which will be used as a reference.

This work is organised as follows: in Sec.~\ref{sec:model_introduction} we present the stochastic model that will be used throughout this manuscript, detail the underlying hypotheses, and introduce relevant observables. In Sec.~\ref{sec:free_pop} we derive the key results for the evolution of individuals without population control. In Sec.~\ref{sec:population_control} we derive the key results for a collection of individuals under different sorts of population control. Conclusions will be finally drawn in Sec.~\ref{sec:conclusion}. In order to improve readability, cumbersome calculations will be relegated to the Appendix.

\section{A simple stochastic model for nuclear reactor physics}
\label{sec:model_introduction}

To fix the ideas and the vocabulary, in the following we
choose to work with a nuclear reactor model, but most
of the results
presented in this paper would apply also to epidemics.

\begin{figure}[t]
    \centering
    \includegraphics[width=\linewidth]{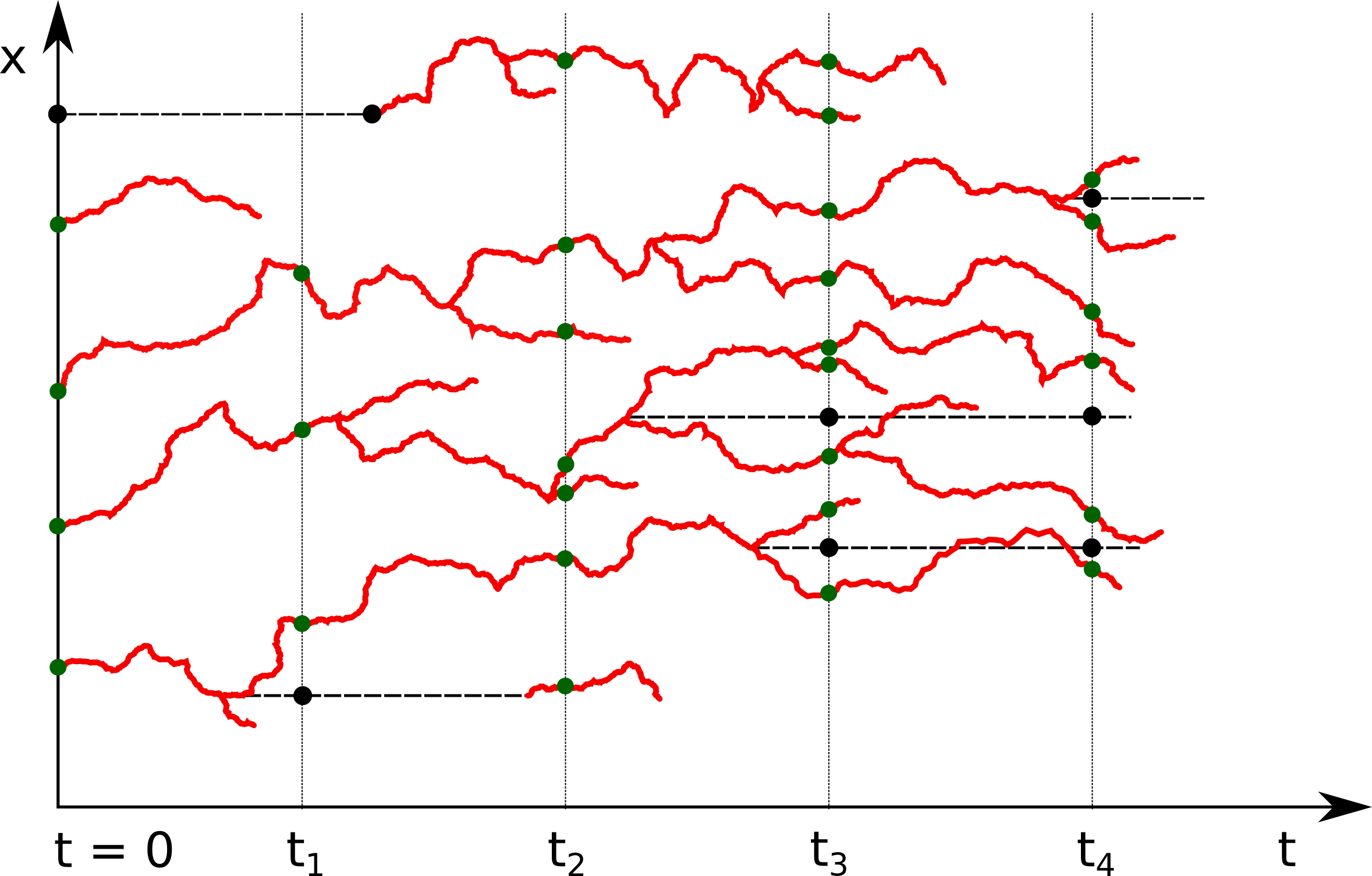}
    \caption{Schematic illustration of the evolution of neutrons and precursors in the simplest model (anarchic model). At time $t=0$, 4 neutrons and 1 precursor are present. Neutrons (red solid lines) diffuse and undergo fission and capture; precursors (dashed black lines) do not diffuse. When they decay, they are replaced by a neutron at the same position. Because of the birth-death process, the total number of neutrons and the total number of precursors fluctuate.} 
    \label{fig:anarch_drawing}
\end{figure}

In the simplest incarnation of our nuclear reactor model
\cite{houchmandzadehNeutronFluctuationsImportance2015}, we represent the
random movement of neutrons as Brownian motion with a diffusion coefficient
${\cal D}$, and we use a Galton-Watson
birth-death process to describe fission and absorption. Each neutron undergoes sterile capture at a rate
$\gamma$ and fission events at a rate $\beta$. Captured neutrons are simply killed. When fission occurs, a
random number of new neutrons are emitted at the position of the incident
neutron, which is killed
\cite{williamsRandomProcessesNuclear1974}. Part of the fission neutrons are emitted
instantaneously and are known as \emph{prompt} neutrons: the probability
to emit $k$ prompt neutrons is denoted as $p_k$. \emph{Delayed} neutrons may further
appear at the spatial site of a fission event after a random, exponentially distributed time with rate $\lambda$,
corresponding to the nuclear decay of certain excited fission fragments, known as \emph{delayed
neutron precursors} \cite{williamsRandomProcessesNuclear1974}.
We denote by $q_k$ the probability that $k$ delayed neutron
precursors are created at a fission event. For illustration, some representative
neutron and precursor histories are shown in Fig.~\ref{fig:anarch_drawing}.
In nuclear reactors, the average precursor decay time $1/\lambda$ is typically
much larger than the average neutron lifetime $1/(\gamma+\beta)$. Furthermore,
denoting respectively by $\nu_{p,1} = \sum_k k p_k$ and $\nu_{d,1} = \sum_k k
q_k$ the average number of prompt and delayed neutrons resulting from fission,
we typically have $\nu_{p,1} \gg \nu_{d,1} $
\cite{williamsRandomProcessesNuclear1974}. For reasons that will become
clear in the following, we are using the notation $\nu_{p,l}$ and
$\nu_{d,l}$ to indicate the
$l$-th falling factorial
moments of the number of neutrons and precursors produced from fission,
respectively, with
\begin{equation}
    \nu_{p,l} = \sum_{k=0}^\infty k (k - 1) \ldots (k-l+1)\,p_k
    \label{eq:prompt_yield}
\end{equation}
and 
\begin{equation}
    \nu_{d,l} = \sum_{k=0}^\infty k (k - 1) \ldots (k-l+1)\,q_k
    \text.
    \label{eq:delayed_yield}
\end{equation}

For the sake of simplicity, our model neglects the energy dependence of the
neutron-matter interaction rates. Furthermore, the
spatial structure
of the reactor core will be simplified by assuming that all the material
properties (diffusion coefficient, fission and capture rates)
are homogeneous. All the physical parameters are taken to be uniform in space
and constant in time. Contrary to Ref.~\citenum{houchmandzadehNeutronFluctuationsImportance2015},
which considered unbounded systems, here the reactor will be modeled
as a finite-size box with
reflection (Neumann) boundary conditions, mimicking the fact that in nuclear
reactors neutron leakage from the core is deliberately kept small. To simplify
matters even
further, we will focus on one-dimensional systems, which are more easily
amenable to analytical solutions: in what follows, we will thus consider a
one-dimensional reactor of half-size $L$, i.e.\ the $[-L,L]$ segment.
This model captures the key physical mechanisms that are responsible
for the fluctuations of the fission chains. 

We will use the parameter values given in Table~\ref{tab:params_table}
throughout this work.

\begin{table}
   \begin{ruledtabular}
    \centering
    \begin{tabular}{cccc}
        & $\theta = 1$ & $\theta = 10^{-1}$ & $\theta = 10^{-3}$\\
        \hline \hline
         \rule{0pt}{3ex}$N$ & $10^2$ & $10^2$ & $10^2$ \\
        $M$ & $10^2$ & $10^3$ & $10^5$ \\
        $L$ & $1$ & $1$ & $1$ \\
        ${\cal D}$ & $10^{-2}$ & $10^{-2}$ & $10^{-2}$ \\
        $\beta$ & $0.2$ & $0.2$ & $0.2$ \\
        $\gamma$ & $0.3$ & $0.3$ & $0.3$ \\
        $\lambda$ & $10^{-1}$ & $10^{-2}$ & $10^{-4}$ \\
        $\nu_{p,1}$ & $2$ & $2$ & $2$ \\
        $\nu_{d,1}$ & $0.5$ & $0.5$ & $0.5$ \\
        $\nu_{p,2}$ & $2$ & $2$ & $2$ \\
        $\nu_{d,2}$ & $0$ & $0$ & $0$ \\
    \end{tabular}
    \caption{Different sets of parameters used in this work. Here $N$ is the
    initial (average) number of neutrons, $M$ is the initial (average) number of precursors, $L$ is the box half-size, ${\cal D}$ is the diffusion
    coefficient, $\beta$ is the fission rate, $\gamma$ is the capture rate,
    $\lambda$ is the decay rate of precursors, $\nu_{p,l}$ and $\nu_{d,l}$ are the
    $l$-th falling factorial moments of the number of neutrons and precursors
    produced from fission, respectively, and $\theta=\lambda/(\beta\nu_{d,1})$.}
    \label{tab:params_table}
    \end{ruledtabular}
\end{table}

\subsection{The critical regime}

Nuclear reactors are operated at the critical regime, where the equilibrium
between births by fission and deaths by capture allows the neutron population
and hence the heat production to be stationary on average \cite{williamsRandomProcessesNuclear1974}. In our model,
criticality is imposed by requiring that 
\begin{equation}
    \beta(\nu_{p,1} + \nu_{d,1}-1) = \gamma \text,
\end{equation}
which is equivalent to equating the production and disappearance rates. The criticality condition does not depend on the number
of neutrons present in the system: indeed, nuclear reactors can be operated at 
steady-state at virtually any power level, the power being proportional to the 
average number of neutrons in the core. Due to
its inherently stochastic nature, the neutron population density
at a spatial site will
display fluctuations (often dubbed ``neutron noise'' \cite{williamsRandomProcessesNuclear1974}) because of random displacements
in and out the spatial cell and because of random changes in the number of
particles following fission and capture events within the cell.

\subsection{Observables of interest}

Fluctuations are characterized by the statistical moments of the neutron
population as a function of position and time. Let $V_1$ and $V_2$ be two
non-overlapping detectors located within the reactor. We are interested
in computing the average number of neutrons $\mathbb{E}[n_i](t_i)$ and precursors
$\mathbb{E}[m_i](t_i)$ present at time $t_i$ in detector $i \in \{ 1,2
\}$, and the two-point correlations between particles detected in $V_1$ at
time $t_1$ and particles detected in $V_2$ at time $t_2$, which stem from the
cross-moments $\mathbb{E}[n_1 n_2](t_1, t_2)$, $\mathbb{E}[n_1 m_2](t_1, t_2)$,
and $\mathbb{E}[m_1 m_2](t_1, t_2)$. Often, it is
preferable to work with continuous observables: we therefore define the neutron
and precursor densities by taking the limits
\begin{align}
     n(x_i,t_i) &=\lim_{V_i \to 0} \frac{\mathbb{E}[n_i](t_i)}{V_i} \\
     m(x_i,t_i) &=\lim_{V_i \to 0} \frac{\mathbb{E}[m_i](t_i)}{V_i},
\end{align}
where $x_i$ is the center of the detector location, $i \in \{ 1,2 \}$. The average population sizes are then obtained by integrating the particle densities over the entire system:
\begin{align}
     n(t) &= \int n(x,t)\,dx \\
     m(t) &=  \int m(x,t)\, dx.
\end{align}
Similarly, we define the pair correlation functions by taking the limits
\begin{subequations}
    \begin{align}
        \begin{split}
            u(x_1,t_1,x_2,t_2) = \lim_{V_1, V_2 \to 0} \frac{\mathbb{E}[n_1 n_2](t_1,t_2)  }{V_1 V_2}
        \end{split}\\
        \begin{split}
            v(x_1,t_1,x_2,t_2) = \lim_{V_1, V_2 \to 0} \frac{\mathbb{E}[n_1 m_2](t_1,t_2)  }{V_1 V_2}
        \end{split}\\
        \begin{split}
            w(x_1,t_1,x_2,t_2) = \lim_{V_1, V_2 \to 0} \frac{\mathbb{E}[m_1 m_2](t_1,t_2)  }{V_1 V_2}
            \text.
        \end{split}%
    \end{align}%
\end{subequations}%
The function $ u(x_1,t_1,x_2,t_2)$ fully characterizes the behavior of
fluctuations and correlations of the neutron population, and in this
respect is key to the investigation of neutron noise. Similarly as for the case of the average densities,
integral quantities can be obtained by integrating the previous equations over the whole spatial domain:
for the two-point neutron correlation function, e.g., we would have
\begin{align}
     u(t_1,t_2) &= \iint u(x,t_1,y,t_2)\, dx\, dy.
\end{align}
The information
content of the two-point correlation function can be
conveniently condensed in the \emph{mean-squared neutron pair distance} $\langle
r^2\rangle(t)$, which is defined as
\begin{align}
    \langle r^2 \rangle(t) = \frac{\displaystyle\iint (x_1-x_2)^2 u(x_1,t,x_2,t)\,dx_1\,dx_2 }{u(t,t)}
    \text,
    \label{eq:def_msd}
\end{align}
which is also helpful in assessing the behaviour of neutron noise
\cite{mulatierCriticalCatastropheRevisited2015, meyerClusteringIndependentlyDiffusing1996}.
In the following we will
characterize the behaviour of these moments under different assumptions concerning
the constraints imposed to the particle population.

\section{Anarchic model}
\label{sec:free_pop}
In the simplest incarnation of our model, we assume that the neutron and
precursor populations are free to evolve according to the stochastic rules
described in Sec.~\ref{sec:free_pop}, without any
constraint. We
will call this system the \emph{anarchic} model (for illustration, see
Fig.~\ref{fig:anarch_drawing}).
The equations for the
particle density and for the two-point correlation functions can be conveniently
established
using the backward master equation approach proposed by P\'al and Bell (which is
special case of Feynman-Kac backward formalism) \cite{williamsRandomProcessesNuclear1974,
pazsitNeutronFluctuationsTreatise2007}.
The idea is to first derive the equations for the moments $n'_{i}(x_0,t_0) = n(x_i,t_i|x_0,t_0)$,
with $i \in \{ 1, 2 \}$, and $u'_{1,2}(x_0,t_0) = u(x_1,t_1,x_2,t_2|x_0,t_0)$
conditioned to having a single initial neutron
starting from position $x_0$ at time $t_0$, treating $x_1$, $x_2$, $t_1$ and $t_2$ as
parameters, and $x_0$ and $t_0$ as variables. The neutron density $n'_i(x_0,t_0)$ is found to satisfy
\begin{equation}
{\cal L}^\dag n'_i(x_0,t_0) = 0,
\label{eq:pal_bell_ave}
\end{equation}
where we have introduced the adjoint linear operator
\begin{multline}
{\cal L}^\dag f(x_0,t_0) =
\frac{\partial f}{\partial t_0} + {\cal D} \nabla^2_{x_0}f +
\alpha_p\,f \\
+ \beta \nu_{d,1} \lambda \int^{\infty}_{t_0} f(x_0, t')\, e^{-\lambda(t'-t_0)}\, dt',
\label{adj_generator_def}
\end{multline}
with the shorthand
\begin{equation}
    \alpha_p = \beta \, (\nu_{p,1} - 1) - \gamma.
\end{equation}
As for the one-particle pair correlation function, we have
\begin{align}
& {\cal L}^\dag u'_{1,2}(x_0,t_0) = - \beta \nu_{p,2} \,n'_1(x_0,t_0) \, n'_2(x_0,t_0) \nonumber\\
& -\beta \nu_{p,1} \nu_{d,1} \lambda\,n_1(x_0,t_0)  \int^{t_2}_{t_0} e^{-\lambda(t'-t_0)} n'_2(x_0,t') dt' \nonumber\\
& -\beta\nu_{p,1} \nu_{d,1}  \lambda\, n_2(x_0,t_0) \int^{t_1}_{t_0} e^{-\lambda(t'-t_0)} n'_1(x_0,t') dt' \nonumber \\
& -\beta\nu_{d,2} \lambda^2 \int^{t_1}_{t_0} \!\!\!\int^{t_2}_{t_0}\!\!\!  e^{-\lambda(t' + t'' -2 t_0)} n'_1(x_0,t') \, n'_2(x_0,t'') dt'  dt''.
\label{eq:pal_bell_cov}
\end{align}
Equations \eqref{eq:pal_bell_ave} and \eqref{eq:pal_bell_cov} are linear and can be both
solved in terms of the Green's function ${\cal G}(x,t|x_0,t_0)$ associated to ${\cal L}^\dag$.
By separation of variables, the Green's function can be expressed as
\begin{equation}
    {\cal G}(x,t,x_0,t_0) = \sum^{+\infty}_{k=0} \varphi_k(x)\varphi_k^{\dagger}(x_0) T_k(t|t_0),
    \label{eqn:series_G}
\end{equation}
where the time-eigenfunctions $T_k(t|t_0)$ read
\begin{multline}
T_k(t|t_0) =
{\left(\omega_k^{+} - \omega_k^{-}\right)}^{-1}\times\\
\left[e^{\omega_k^{+} (t-t_0)} (\omega_k^{+}+\lambda)- e^{\omega_k^{-} (t-t_0)} (\omega_k^{-}+\lambda)\right]
\text,
\label{eqn:temporal_mode}
\end{multline}
with two associated families of eigenvalues
\begin{equation}
\omega_{k}^\pm =\frac{\alpha_k + \alpha_p -\lambda \pm \sqrt{(\alpha_k + \alpha_p +\lambda)^2+4\lambda \beta \nu_{d,1}}}{2}.
\label{eq:omega_k}
\end{equation}
The quantities $\alpha_k$ are the eigenvalues
associated to operator ${\cal
D}\nabla^2_x$, with the corresponding eigenfunctions $\varphi_k(x)$. For
reflection boundary conditions, we have $\alpha_k = - {\cal D} ( k \pi/2 L)^2$ and
\begin{equation}
    \varphi_k(x) = \cos\left( \frac{k \pi x}{2 L} \right)
    \label{eq:phi}
\end{equation}
for $ k \ge 0$.
The functions $\varphi_k^{\dagger}(x_0)$ are obtained from
the orthonormality condition of the eigenfunctions, and for reflection boundary
conditions read 
\begin{equation}
    \varphi^\dag_k(x_0) =
    \begin{cases}
      \displaystyle\frac{1}{2 L} & \text{if } k = 0 \\
      \displaystyle\frac{1}{L}\cos\left( \frac{k \pi x_0}{2 L}\right) & \text{if } k \ge 1.
    \end{cases}
    \label{eq:phi_dagger}
\end{equation}
In nuclear reactors, the two families $\omega_{k}^\pm$ are
widely separated because of the typical values of $\lambda$,
$\alpha_p$ and $\beta \nu_{d,1}$. Loosely speaking, the $\omega_k^+$ family is associated to
the dynamics of precursors, while $\omega_k^-$ describes prompt dynamics.

Once the single-particle moments are known, it is then straightforward to
generalize the framework to describe a collection of $\mathcal{N}$ \emph{independent and
identically distributed} individuals. The details of the
calculations can be found in Appendix~\ref{appendix:backward_formalism}.
In particular, if we denote the single-particle
source distribution as $\mathcal{Q}(x_0,t_0)$, the density $n(x,t)$ reads
\begin{equation}
 n(x,t) = {\cal N} \iint_{-\infty}^{t}  {\cal Q}(x_0, t_0)
 {\cal G}(x,t|x_0,t_0)\,dx_0\,dt_0
 \label{eq_ave_density_source}
 \text,
\end{equation}
which expresses a linear superposition principle. The pair correlation function
can be also explicitly computed based on the Green's function, and reads 
\begin{multline}
u(x_1,t_1,x_2,t_2) = \\
\frac{{\cal N}-1}{{\cal N}} n(x_1,t_m) \, n(x_2,t_M)
+ {\cal G}(x_2,t_M|x_1,t_m) \, n(x_1,t_m)\\
+\beta \nu_{p,2} \int_{0}^{t_m} \!\!\! \int {\cal G}(x_1, t_m|x,t) \, {\cal G}(x_2,t_M|x,t)\, n(x,t)\mathit{dx}\mathit{dt}\\
+\beta \nu_{p,1} \nu_{d,1} \int_{0}^{t_m}\!\!\! \int {\cal G}(x_1,t_m|x,t)\, c(x_2,t_M|x,t) \,n(x,t)\mathit{dx}\mathit{dt} \\
+ \beta\nu_{p,1} \nu_{d,1} \int_{0}^{t_m}\!\!\! \int {\cal G}(x_2,t_M|x,t)\, c(x_1,t_m|x,t) \,n(x,t)\mathit{dx}\mathit{dt}\\
+ \beta\nu_{d,2} \int_{0}^{t_m}\!\!\! \int c(x_1,t_m|x,t) \,c(x_2,t_M|x,t) n(x,t)\mathit{dx}\mathit{dt} \text,
\label{eq_def_corr}
\end{multline}
where $t_m = \min\{t_1,t_2\}$, $t_M=\max\{t_1,t_2\}$, and
the quantity 
\begin{equation}
    c(x_1,t_1|x,t) = \lambda \int_{t}^{t_1} {\cal G}(x_1,t_1|x,t') e^{-\lambda(t'-t_1)}\,dt'
    \label{eq:green_precursors}
\end{equation}
represents the expected number of neutrons detected at $(x_1,t_1)$ produced by a
precursor at $(x,t)$, with $t_1 > t$.

The terms appearing in the correlation function in Eq.~\eqref{eq_def_corr} can
be given a physical interpretation. The first term describes the contribution of
the detection of 
uncorrelated particles, i.e.\ particles that do not share any common ancestor.
The 
second term describes self-correlations, i.e.\ the detection of the same
particle history 
at $(x_1,t_m)$ and later at $(x_2,t_M)$. The four remaining
terms represent the possible 
contributions due to fission events: the first one corresponds to the
correlations induced by detection at $(x_1,t_M)$ and $(x_2,t_M)$ of neutrons
stemming from different prompt neutrons produced
by a common fission event at $(x,t)$; the following two terms are symmetrical 
and correspond to correlations induced by detection at $(x_1,t_M)$ and
$(x_2,t_M)$ of neutrons stemming from a prompt neutron and a precursor produced
by a common fission event at $(x,t)$; finally, the remaining term 
corresponds to correlations induced by detection at $(x_1,t_M)$ and $(x_2,t_M)$
of neutrons stemming from different precursors produced
by a common fission event at $(x,t)$.

The reactor is assumed to be in the critical regime. It is then possible to choose
the single-particle source ${\cal Q}(x_0, t_0)$ so that the average neutron
density $n(x,t)$ is stationary at any time $t>0$: this is for instance achieved by taking
${\cal Q}(x_0, t_0) = {\cal Q}_c(x_0, t_0)$, with
\begin{align}
    {\cal Q}_c(x_0, t_0) &= \frac{1}{2L} \left[\frac{\theta}{1+\theta}\delta(t_0)+\frac{1}{1+\theta}\lambda \exp(-\lambda t_0)\right]
                       \text,
    \label{eq:source_distribution}
\end{align}
where $\theta = \lambda / (\beta \nu_{d,1})$. Typically, for nuclear
reactors, $\theta \sim 10^{-4}$ to $10^{-5}$. This particular choice of ${\cal
Q}(x_0, t_0)$ will be called the \emph{critical source}. The first
term in
Eq.~\eqref{eq:source_distribution} corresponds to neutrons
appearing at
$t_0 = 0$ with probability $\theta/(1+\theta)$, while the second term corresponds to delayed neutrons
appearing at
exponentially distributed times from the decay of the precursors initially
present at $t_0 = 0$, with complementary probability $1/(1+\theta)$.
For our
simple homogeneous one-dimensional reactor with reflection boundary conditions, 
the spatial distribution $1/(2L)$
is flat over the domain (Eq.~\eqref{eq:phi}).

\subsection{Analysis of the pair correlation function}

\begin{figure}[t]
    \centering
    \includegraphics[width=\linewidth]{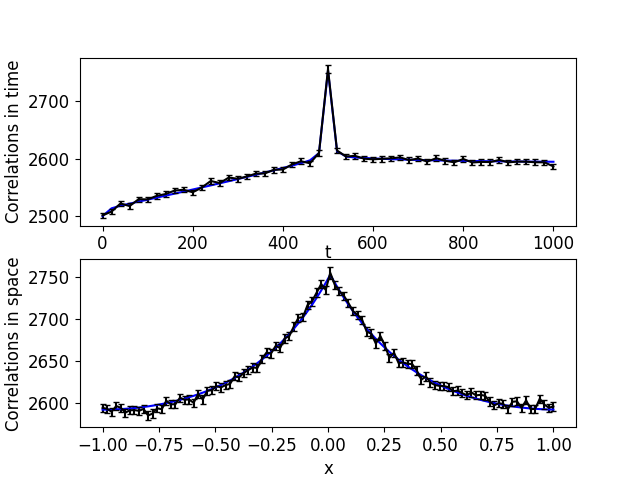}
    \caption{Pair correlation function for a critical population starting with, on average, $N =
    100$ prompt neutrons and $M = 10^3$ precursors, for a deterministic total population size ${\cal N}
    = N + M$. Parameters are taken as in Table~\ref{tab:params_table} for $\theta = 10^{-1}$. On the graph we plot $u(x_1,t_1,x_2,t)$ for $x_1 = x_2 = 0$ and $t_1 = T/2$. The bottom graph shows $u(x_1,t_1,x,t_2)$ the spatial shape when $t_1 = t_2 = T$ and $x_1 = 0$. Blue: analytical results from Eq.~\eqref{eq_def_corr}.
    Black: Monte Carlo simulations result with $10^6$ replicas.}
    \label{fig:h_anarch_ref}
\end{figure}

\begin{figure}[t]
    \centering
    \includegraphics[width=\linewidth]{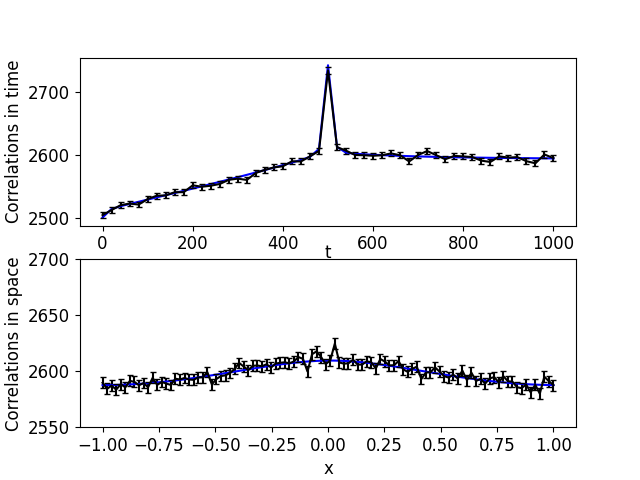}
    \caption{Same as Fig.~\ref{fig:h_anarch_ref}, except that the top graph gives the time dependency for $x = 0$ and $x_2 = -0.1$. The bottom graph shows spatial correlations with $t_1 = T = 990$ and $t_2 = 1000$.}
    \label{fig:h_anarch_ref_diff}
\end{figure}

Inspection of Eq.~\eqref{eq_ave_density_source} shows that the average neutron density
in a critical reactor starting form the critical source in Eq.~\eqref{eq:source_distribution}
is trivially constant, as expected,
and reads $n(x,t) = N/(2L)$. This value depends on the number $N$ of neutrons
emitted from the source and on the reactor size $2L$. Correspondingly, the average total
number of neutrons $n(t)$ will be constant and equal to $N$, and it can be shown that
the average total number of precursors $m(t)$ will be also constant and equal to $M =
N/\theta$.

The correlation function in Eq.~\eqref{eq_def_corr}, on the contrary, is not stationary
for a critical system. While Eq.~\eqref{eq_def_corr}
is exact, the evaluation of ${\cal G}$ requires the
numerical computation of an infinite series. In
the following, the semi-analytical
solution is henceforth obtained using
Eqs.~(\ref{eq:first_anarch_analytical}-\ref{eq:last_anarch_analytical}) by
summing the series up to order $k=1000$. The resulting correlation function
has been carefully verified against Monte Carlo simulations, as shown in
Figs.~\ref{fig:h_anarch_ref} and \ref{fig:h_anarch_ref_diff}. It is more
convenient to plot the \emph{corrected pair correlation function}
$\Tilde{u}(x_1,t_1,x_2,t_2) = u(x_1,t_1,x_2,t_2) - \delta(x_1-x_2)\, n(x_1,t_1)$,
which is
equivalent to ignoring the singular term $\delta(x_1-x_2)$ due to the self-correlations
occurring in the system at $t_1 = t_2$. Note that this does not remove all of the
self-correlation contributions, as self-correlations also include contributions from
points $x_1 \neq x_2$.
By abuse
of language, and when it does not impede understanding, in the following we shall
call $\Tilde{u}$ the ``pair correlation function'' nonetheless. Since the function has four
independent variables, for the sake of simplicity
we present one-dimensional cuts for fixed values of the other variables.

Figure~\ref{fig:h_anarch_ref} shows the time shape of the correlation function 
when particles are detected at the same spatial position $x_1=x_2$ (top panel), and the
spatial shape when the detection times $t_1=t_2$ are the same (bottom panel). The
detection positions chosen for Fig.~\ref{fig:h_anarch_ref} are far from
the boundaries of the reactor. Figure~\ref{fig:h_anarch_ref_diff} illustrates
the time shape of the correlation function for two different
detection positions (top),
and its spatial shape for two different detection times (bottom).

Figures~\ref{fig:h_anarch_ref} (top panel) and \ref{fig:h_anarch_ref_diff} (top
panel) show that, after an initial transient, $t\lesssim50$, the time profile of the
pair correlation function exhibits a linear build-up.
For $t_1 > t_2$,
correlations saturate. Indeed, the linear build-up describes fissions happening at time
$t_1$ that may contribute to further fissions occurring at a fixed $t_2$, meaning that
obviously fissions occurring at $t_1 > t_2$ cannot contribute.

\begin{figure}
     \centering
     \includegraphics[width=\linewidth]{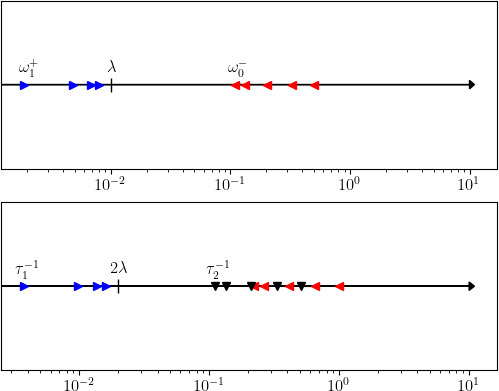}
     \caption{Scatter plot of the characteristic time constants of the Green's function (top) and of the one-time, two-point correlation function (bottom), for the anarchic system, up to order $k=5$. Top: blue right-pointing triangles represent $|\omega_k^{+}|$; red left-pointing triangles represent $|\omega_k^{-}|$. Bottom: blue right-pointing triangles represent $|2 \omega_k^{+}|$; black down-pointing triangles represent $|\omega_k^{+} + \omega_k^{-}|$; red left-pointing triangles represent $|2 \omega_k^{-}|$. Note that $\omega_0^-=0$ and $2\omega_0^-=0$ cannot be represented in either plot (the system is critical). The dominant mode for each family ($k=0$) is the leftmost one. The parameter values are the same as in Fig.~\ref{fig:h_anarch_ref}.}
     \label{fig:anarch_modes}
\end{figure}

The analysis of
$u(x_1,t_1,x_2,t_2)$ shows that the correlation function displays several typical time
scales. First, the spatial boundedness of the
system introduces time scales that are related to diffusion. In the absence of
delayed neutrons,
the relaxation of the space profile of the
correlation function is characterized by several time scales. The prompt time
scales accumulate around and are
asymptotically dominated by the mixing time, which is related to the time a neutron needs in order
to explore the whole physical space \cite{zoiaClusteringBranchingBrownian2014}.
With the introduction of precursors, additional remarkable time scales appear.
As shown in Eq.~\eqref{eq_def_corr}, the two-point, two-time correlation function involves
products of the Green's function, Eq.~\eqref{eqn:series_G}, which has the structure of an
infinite sum of exponential modes with characteristic time constants $\omega_k^+$ and
$\omega_k^-$. The correlation function can then be expressed as an infinite sum of
exponential modes with characteristic time constants $2 \omega_k^+$, $2 \omega_k^-$,
$\omega_k^{+} + \omega_k^{-}$ and $\omega_k^{+} - \omega_k^{-}$ (see
Appendix~\ref{appendix:anarch_sol} for details). If we restrict our attention to the
one-time, two-point correlation function, $u(x_1,x_2,t)=u(x_1,t,x_2,t)$, then the $\omega_k^{+} -
\omega_k^{-}$ modes disappear. The first few characteristic time
constants for each family are represented in Fig.~\ref{fig:anarch_modes}. We can
identify four different interesting time scales:
\begin{itemize}
    \item the separation of the first two characteristic time constants of the $\omega_k^+ +
    \omega_k^-$ family is given by $(\omega_0^+ + \omega_0^-) - (\omega_1^+ + \omega_1^-)
    = -\alpha_1 = 4 L^2/ {\cal D} \pi^2 = \tau_D^{-1}$. This is exactly the inverse of the mixing
    time for a system without delayed neutrons, as studied in
    Ref.~\citenum{zoiaClusteringBranchingBrownian2014}. For the
    parameter values of Figs.~\ref{fig:h_anarch_ref} and \ref{fig:h_anarch_ref_diff}, the prompt mixing
    time would be $\tau_D\simeq41$.

    \item on a time scale of the order of $\tau_2 = |\omega_0^+ + \omega_0^-|^{-1} =
    (\beta \nu_{d,1} + \lambda)^{-1}$, the correlation function reaches a quasi-stationary
    regime, after prompt dynamics has stabilized but before delayed dynamics
    comes into play. Note that the expression of $\tau_2$ is independent of the spatial
    characteristics of the
    system, such as its spatial extent or the diffusion coefficient. Therefore, this time
    scale corresponds to a collective behaviour. For times of the order of a few $\tau_2$, the
    prompt modes contribute a time-independent term to the correlation function. The delayed
    modes, on the other hand, have time scales that are much longer than $\tau_2$; therefore, the
    remaining exponentials terms $\exp(2\omega_k^+ t)$ can be replaced with their first-order
    Maclaurin development, yielding a linear dependency on time. This explains the presence of
    linear build-up regimes in Figs.~\ref{fig:h_anarch_ref} and \ref{fig:h_anarch_ref_diff}. For
    the parameters of these figures, we have $\tau_2 \simeq 9$.

    \item on a time scale of the order of $\tau_1 = |2\omega_1^+|^{-1}$, the correlation
    function acquires its asymptotic spatial shape. In this sense, this time scale is
    analogous to the mixing time of prompt systems; however, given the expression of
    Eq.~\eqref{eq:omega_k}, it involves the precursor decay constant $\lambda$ and is
    therefore generally much longer than the prompt mixing time, which is given by
    $\tau_D$. For the parameter values of Figs.~\ref{fig:h_anarch_ref} and
    \ref{fig:h_anarch_ref_diff}, we have $\tau_1\simeq 269$.
    
    \item for times much longer than $\tau_1$, all the exponential terms in the correlation
    function have died out. The spatial shape of the correlation function is frozen, but the
    overall scale factor increases linearly with $t$. It has been previously shown by
    Houchmandzadeh et al.\ 
    \cite{houchmandzadehNeutronFluctuationsImportance2015} (and our calculations confirm this)
    that the slope of this eventual linear regime is $\tau_E^{-1}$, where
    \begin{equation}
        \tau_E = \frac{N(1+\theta)^2}{\beta \nu_2 \theta^2}
        \text,
        \label{eq:extinction_time_anarchic}
    \end{equation}
    where
    \begin{equation*}
        \nu_2 = \nu_{p,2} + 2 \nu_{d,1} \nu_{p,1} + \nu_{d,2}\text,
    \end{equation*}
    and $N = {\cal N}\theta / (1+\theta)$ is the average number of neutrons in the system. After a time of the order of $\tau_E$, the standard deviation of the total population size
    becomes equal to the mean population size. This is the regime of the critical catastrophe:
    because of the unbounded fluctuations, the population will almost surely go eventually
    extinct, despite the fact that the average population size is constant
    \cite{williamsRandomProcessesNuclear1974, pazsitNeutronFluctuationsTreatise2007}.
    This apparently paradoxical behaviour is a well-known characteristic of critical
    birth-and-death processes. The quantity $\tau_E$ thus physically represents the typical
    \emph{extinction time} of the system. For the parameter values of Figs.~\ref{fig:h_anarch_ref}
    and \ref{fig:h_anarch_ref_diff}, the extinction time is $\tau_E\simeq 1.5\times 10^4$;
    therefore, this regime is not observable in the figures above.
\end{itemize}

The space profile of the correlation function, shown in Fig.~\ref{fig:h_anarch_ref}
(bottom panel), has the tent-like shape that is typical of
clustering behaviour. Despite the fact that the mean population density is uniform
and constant, specific histories can exhibit strong patchiness (clusters) where
pairs of particles will tend to be located at short distances from each other. This
is coherent with
previous findings for a reactor model neglecting delayed neutrons
\cite{dumonteilParticleClusteringMonte2014, zoiaClusteringBranchingBrownian2014,
zoiaNeutronClusteringSpatial2017}. Figure~\ref{fig:h_anarch_ref_diff} (bottom)
suggests a relaxation toward a flat distribution for $t_1$ and $t_2$ sufficiently
far from each another. Further analyses (not shown here for conciseness) confirm
this conjecture.

It is tempting to characterize the importance of neutron clustering by defining
a dimensionless parameter $\eta = \tau_1 / \tau_E$, i.e.\ the ratio of the two longest time scales
in the system. When $\eta \gg 1$, clustering will relax slowly over a time of the order of the
extinction time, and therefore it will be observable over the whole lifetime of the system.
It is interesting to consider how the dimensionless parameter $\eta$ scales with
$\theta$ for small $\theta$, which is the regime of interest for nuclear reactors. By
replacing $\lambda=\theta \beta \nu_{d,1}$ in Eq.~\eqref{eq:omega_k} and developing in
Maclaurin series in $\theta$, we obtain
\begin{equation}
    \omega_1^+ \simeq
    \frac{\theta}{\alpha_1^{-1} + \alpha_p^{-1}}
    \text.
\end{equation}
Hence,
\begin{equation}
    \tau_1\simeq\frac{\alpha_1^{-1} + \alpha_p^{-1}}{2\theta}
    \text.
    \label{eq:tau1_maclaurin_theta}
\end{equation}
This should be contrasted with the mixing time for a system without delayed
neutrons, which is $\tau_D=1/\alpha_1$. Even in the regime where diffusion
dominates ($\alpha_1\ll \alpha_p$), we can see that the presence of delayed
neutrons slows down mixing by a factor of $1/\theta$. Unsurprisingly, this is
the factor that is introduced by delayed neutrons in the characteristic
response times of the \emph{mean} reactor populations around criticality
\cite{houchmandzadehNeutronFluctuationsImportance2015}. Thus, mixing is
essentially a phenomenon that characterizes the behaviour of the average
neutron and precursor densities.

When Eq.~\eqref{eq:tau1_maclaurin_theta} is replaced in the definition of the clustering
parameter $\eta$, we find
\begin{equation}
    \eta = \frac{\tau_1}{\tau_E} \simeq \frac{\alpha_1^{-1} + \alpha_p^{-1}}{2N}\beta \nu_2
    \theta
    \text.
\end{equation}
This goes to show that the presence of delayed neutrons suppresses clustering, because
of the different scaling of $\tau_1$ and $\tau_E$ with $\theta$. It
is worth remarking in passing that the fact that the extinction time scales as
$\theta^{-2}$ for small values of $\theta$ is not a coincidence, but it is due to the
fact that $\tau_E$ is a purely stochastic quantity, which can only be defined in terms
of the second-order moments of the populations.


\subsection{Analysis of the mean-squared pair distance}
\label{section:anarch_cluster_dist}

The quantity $\langle r^2\rangle (t)$ yields the average square distance
between pairs of particles (observed at the same time), and in this respect provides
information on their tendency toward clustering
\cite{mulatierCriticalCatastropheRevisited2015, meyerClusteringIndependentlyDiffusing1996}.
At time $t_0=0$, we have
\begin{equation}
    \langle r^2\rangle(0) = \langle r^2\rangle_{iid} = \frac{2 L^2}{3},
    \label{eq:uncorrelated_pair_dist}
\end{equation}
since the neutrons
are independently and uniformly distributed within the reactor.
At later times, neutrons exhibit
correlations induced by fission events and
correspondingly $\langle r^2 \rangle(t) < \langle r^2 \rangle_{iid}$, which is a
signature of the spatial clustering regime.
For the anarchic model, we can obtain
semi-analytical formulas for $\langle r^2\rangle(t)$ in the form of a
series, whose expression can be found in
Eq.~\eqref{eq:anarch_pair_distance}, generalizing previous findings for the model
without delayed neutrons
\cite{mulatierCriticalCatastropheRevisited2015}.

\begin{figure}
    \includegraphics[width=\linewidth]{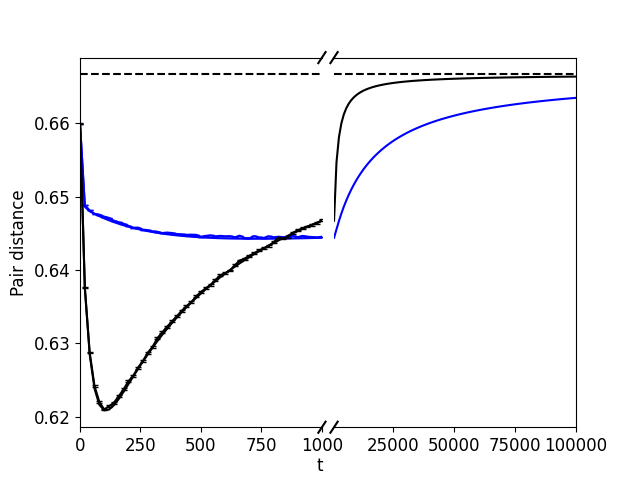}
    \caption{Mean-squared pair distance in a one-dimensional box $[-L,L]$ for critical populations using two sets of parameters taken from Table~\ref{tab:params_table}. Error bars represent Monte Carlo simulation with $10^6$ replicas, full lines the analytical integration of Eqs.~\eqref{eq_def_corr} and \eqref{eq:def_msd}. The blue curves correspond to $\theta = 10^{-2}$, and the black ones to $\theta = 1$.}
    \label{fig:anarch_r2}
\end{figure}

Figure~\ref{fig:anarch_r2} shows the time evolution of the
mean-squared pair distance for a choice of model parameters.
The formula from Eq.~\eqref{eq:anarch_pair_distance} is compared to the result of the
Monte Carlo simulation. Pair distance in Monte Carlo is estimated by taking the ensemble average of the squared distances of all the neutron pairs present in the system at a given time, normalized by the ensemble average of the square of the number of particles at a given time. For times of the order of the extinction time, the
quantity $\langle r^2 \rangle(t)$ asymptotically reverts to the uncorrelated value $\langle r^2
\rangle_{iid}$, with a convergence speed that depends on the model parameters.
This is explained by the fact that for $t
\gg \tau_E$ the population in a critical regime almost certainly dies out and
correlations become spatially flat over the entire reactor. This result is again coherent with previous findings
for the case where precursors were neglected 
\cite{mulatierCriticalCatastropheRevisited2015}.

\section{Models with population control}
\label{sec:population_control}

So far, we have considered the case of freely evolving particle populations.
In nuclear reactors, various feedback mechanisms act against the excursions of
the neutron population. An example of utmost importance is provided by the Doppler
effect. The probability of neutron-matter interaction depends on the temperature of the
medium; in nominal conditions, the probability of sterile capture increases with
increasing temperature \cite{bell_nuclear_1970}. Since the temperature of the
reactor is driven by the number of neutrons in the system, the overall result of the
Doppler effect is that deviations from the mean in the number of neutrons in the reactor
tend to be quenched. The typical time scale of the Doppler effect is much shorter
than the neutron lifetime, so that the feedback acts almost instantaneously on the
behaviour of the fission chains \cite{bell_nuclear_1970}.

For the model without delayed neutrons, these effects have been investigated by
introducing an idealized control mechanism that acts as follows: whenever a neutron is
created by fission, another neutron is randomly
chosen and eliminated \cite{mulatierCriticalCatastropheRevisited2015,
zoiaNeutronClusteringSpatial2017}.
This model was introduced based on heuristic arguments following previous works
in the context of theoretical ecology
\cite{meyerClusteringIndependentlyDiffusing1996,
zhangDiffusionReproductionProcesses1990}: it
preserves exactly the total number of individuals in the system and
has a deep impact on the behaviour of the correlations. In particular, it has been shown
to induce an upper bound on the amplitude of neutron noise
and thus prevent the occurrence of the
critical catastrophe \cite{mulatierCriticalCatastropheRevisited2015,
zoiaNeutronClusteringSpatial2017}.

As a stepping stone towards a precise understanding of 
the effect of a physical feedback on the time and space shape of the
correlation function, in the following we will introduce several simplified
population control models that extend previous attempts by including precursors.

The rigorous derivation of the equations for the models with population control is
provided in the Appendix~\ref{appendix:master_equations}. The enforcement of control mechanisms,
which correlate birth and death events, breaks the independence of particle histories,
so that the backward formalism can hardly be used
\cite{palNeutronFluctuationsMultiplying2006}.
We will resort to a forward master equation approach instead. For the sake of
simplicity, we will focus exclusively on the case of binary fission and on the production of at most one delayed neutron precursor, although our
findings can be straightforwardly extended to general $p_k$ and $q_k$ distributions, at the
expense of more cumbersome formulas. For the same reason, we shall
take $t_1 = t_2 = t$, i.e.\ consider spatial correlations with both particles observed
at the same time.

\subsection{Control of neutrons in a system of neutrons and precursors}
\label{sss:N_cst}

Let us consider a population of neutrons and precursors, evolving based on the
stochastic rules described in
Sec.~\ref{sec:free_pop}. Since only neutrons are affected by physical feedback
mechanisms (and are ultimately responsible for
the physical changes occurring in the reactor), it would be natural to enforce
population control on neutrons alone, which can be achieved as follows.

\begin{figure}[t]
    \centering
    \includegraphics[width=\linewidth]{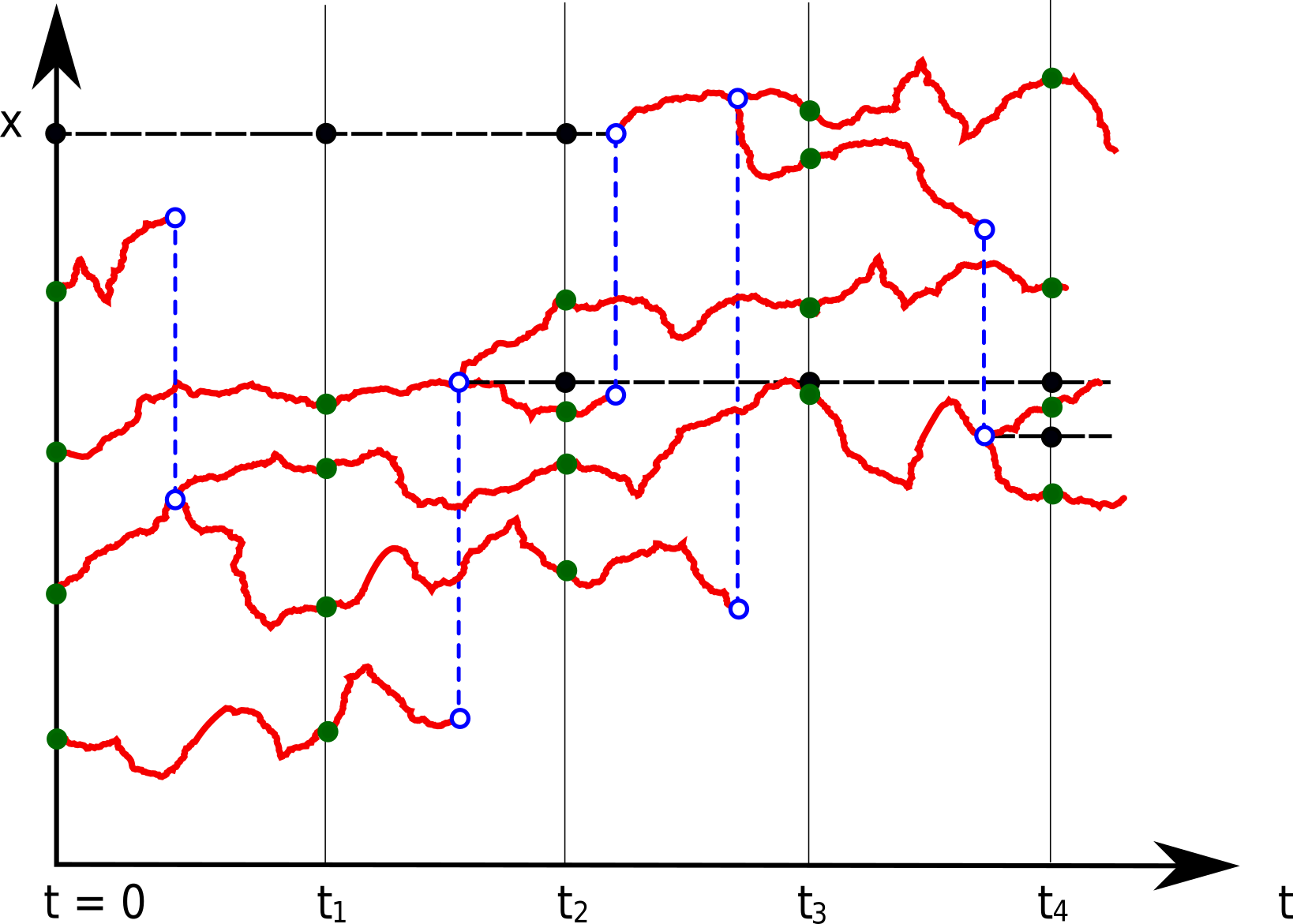}
    \caption{A typical history for a system of 4 neutrons and 1 precursor in the $N$-control model. Neutrons diffuse (in red) until they branch or are destroyed by another neutron being produced in the system (these correlated event are shown in blue dashed lines), and precursor life-time is shown by black dashed-lines.}
    \label{fig:partial_control_drawing}
\end{figure}

Each neutron induces fission with a rate of $\beta$; the fission event produces
exactly two prompt neutrons (with probability $p_2 = 1$).
Additionally, fission events produce a precursor with
probability $q_1$, and no precursor with the complementary probability $q_0 = 1-q_1$.
Upon decay, with rate $\lambda$, the precursor yields a delayed neutron. In this specific case,
we have $q_1 = \nu_{d,1}$, which is not true in general. In order to
ensure a constant number $N$ of neutrons in the system, whenever a neutron is produced,
either from fission or from a decaying precursor, another neutron is randomly chosen and
eliminated (for illustration, see Fig.~\ref{fig:partial_control_drawing}). Observe that the
total number of precursors is left free to fluctuate. We will call this scheme the
\emph{$N$-control} algorithm.

We begin our analysis by addressing the behaviour of the population
sizes. First we denote by $N$ and $M$ the (arbitrary) initial number of
neutrons and precursors, which we assume to be uniformly distributed in
space. From the forward master equation (see
Appendix~\ref{appendix:master_equations} for details), we deduce the equations for the
average neutron and precursor population sizes, which read
\begin{subequations}%
\begin{align}
        \frac{\partial}{\partial t} n(t) &= 0  \\
        \frac{\partial}{\partial t} m(t) &= \beta \nu_{d,1} N - \lambda m(t)
        \text.
\end{align}
\label{eq:eqdiff_mtot_cons_n}
\end{subequations}%
These equations can be readily solved, together with the initial conditions, and yield 
\begin{align*}
  n(t) &= N  \\
  m(t) &= \frac{\beta  \nu_{d,1}}{\lambda}N \left(1-e^{-\lambda t}\right) + M e^{-\lambda t}
  \text.
\end{align*}
For the \emph{$N$-control} model, the average total precursor number $m(t)$ exponentially relaxes to the same equilibrium solution as for the anarchic model, i.e.,
$m_\infty=N \, \beta  \nu_{d,1}/\lambda = N / \theta$. If the initial condition is chosen as $N = M \theta$, we simply obtain $m(t) = M$ for any $t$. The equations for the second-order moments of the total population can be also derived, and read
\begin{subequations}
\begin{align}
  \frac{\partial}{\partial t} u(t) &= 0 \\
  \frac{\partial}{\partial t} v(t) &= \beta  \nu_{d,1} \, u(t) - \lambda \, v(t) \label{eq:eqdiff_vtot_cons_n}\\
  \frac{\partial}{\partial t} w(t) &= \beta  \nu_{d,1} N\, (2 m(t) + 1) - \lambda(2 w(t) - m(t))
  \text.
\end{align}%
\label{eq:eqdiff_uvwtot_cons_n}
\end{subequations}%
The solutions to these equations are
\begin{align}
   u(t) &= N^2 \\
   v(t) &= N m(t) \\
   w(t) &= m(t)^2 + m(t) - M e^{-2\lambda t}
   \text.
\end{align}
In the \emph{$N$-control} model, the variance of the neutron population size is
zero, as expected, which means in particular that the critical catastrophe
is averted. The variance of the precursor population size converges
exponentially to $m_\infty$. It is actually easy to show that the
distribution of the precursor population size is asymptotically Poissonian,
but we omit this proof for the sake of conciseness.

Starting from a fully spatially discretized master equation (see Appendix
\ref{appendix:master_equations}), we can then obtain the equations for the spatial
moments of the populations. The equations for the first moments read
\begin{subequations}
\begin{align}
  \frac{\partial}{\partial t} n(x,t) &= {\cal D} \nabla^2_x n(x,t)  + \lambda \left( m(x,t) -\frac{1}{N} H^{n}(x,t) \right)\\
  \frac{\partial}{\partial t} m(x,t) &= \beta  \nu_{d,1} \, n(x,t) - \lambda \, m(x,t)
  \text,
\end{align}%
  \label{eq:N_cst_1st}%
\end{subequations}%
where we have defined the cross-moment
\begin{equation}
H^{n}(x,t)=\lim_{V_i \to 0} \frac{1}{V_i}\mathbb{E}[n_i(t) m(t)]
\text,
\end{equation}
with $V_i$ centered around $x$. A somewhat surprising fact is that the term $H^n(x,t)$
appearing in the equation for the average is actually a second-order moment: this means
that unfortunately the hierarchy of the spatial moment equations is not closed
and analytical solutions are therefore out of reach for this model.  This is to
be contrasted with the case of the moments of the integral quantities,
Eqs.~\eqref{eq:eqdiff_mtot_cons_n} and \eqref{eq:eqdiff_uvwtot_cons_n}, which
are closed.  On the other hand, Monte Carlo simulations show that, taking $M =
N / \theta$ and a uniform spatial distribution, then the first moments behave
like in the anarchic case and verify
\begin{align}
    n(x,t) &= \frac{N}{2 L} \\
    m(x,t) &= \frac{M}{2 L} \text.
\end{align}
Similarly, the equations for the pair correlation functions
\begin{align*}
    \Tilde{u}(x,y,t) & = u(x,t,y,t) - n(x,t) \delta(x-y)\\
     v(x,y,t) &= v(x,t,y,t) \\
    \Tilde{w}(x,y,t) &= w(x,t,y,t) - m(x,t) \delta(x - y),
\end{align*}
involve third-order moments:
\begin{widetext}
\begin{subequations}
\begin{align}
    \begin{split}
    \frac{\partial}{\partial t} \Tilde{u}(x,y,t) &= {\cal D}\left(\nabla_x^2 +
      \nabla_y^2\right)\Tilde{u}(x,y,t) - \frac{2 \beta}{N - 1} \Tilde{u}(x,y,t) + \lambda
      C_{N} \left( v(x,y,t) + v(y,x,t)\right) - \frac{2\lambda}{N}H^{nn}(x,y,t)\\
      &+ \delta(x-y)\left(2\beta n(x,t) + \frac{2\lambda}{N} H^{n}(x,t) \right)
    \end{split}
    \label{eq:N_cst_2nd_u}\\
    \begin{split}
    \frac{\partial}{\partial t} v(x,y,t) &= {\cal D}\nabla_x^2 v(x,y,t) - \lambda C_{N} v(x,y,t) - \frac{\lambda}{N} H^{nm}(x,y,t) + \beta  \nu_{d,1} C_{N-1} \Tilde{u}(x,y,t) + \lambda \Tilde{w}(x,y,t) \\
    &+ \delta(x-y) \frac{\beta  \nu_{d,1} N}{N-1} n(x,t)
    \end{split}
    \label{eq:N_cst_2nd_v}\\
    \frac{\partial}{\partial t} \Tilde{w}(x,y,t) &= \beta  \nu_{d,1} \left(v(x,y,t) + v(y,x,t)\right) - 2\lambda \Tilde{w}(x,y,t)
    \label{eq:N_cst_2nd_w}
    \text.
\end{align}%
    \label{eq:N_cst_2nd}%
\end{subequations}%
\end{widetext}
For the sake of brevity, we have defined the two third-order
cross-moments
\begin{align*}
  H^{nn}(x,y,t)&=\lim_{\substack{V_i,V_j \to 0}} \frac{1}{V_i \, V_j} \mathbb{E}[n_i(t) n_j(t) m(t)] \\
  H^{nm}(x,y,t)&=\lim_{\substack{V_i,V_j \to 0}} \frac{1}{V_i \, V_j} \mathbb{E}[n_i(t) m_j(t) m(t)]
  \text.
\end{align*}
and the shorthand $C_{N} = (N-1)/N$. The detection volumes $V_i$ and $V_j$ are assumed
to be centered around $x$ and $y$, respectively. It is interesting to notice that we can recognize in
Eq.~\eqref{eq:N_cst_2nd_u}
the \emph{the renewal time} (or, equivalently, the fixation time \cite{SUTTON20171211}) $\tau_{R,0} = (N - 1)/(2\beta)$ for a population
of prompt neutrons under population control without precursors
\cite{meyerClusteringIndependentlyDiffusing1996}. In the absence of delayed neutrons, the renewal time is
the typical time it takes for all surviving neutrons in the system to share the same
common ancestor. Although this definition can be extended to a population of neutrons
and precursors, we obviously expect the associated renewal time to differ from
$\tau_{R,0}$.

\begin{figure}
     \centering
     \includegraphics[width=\linewidth]{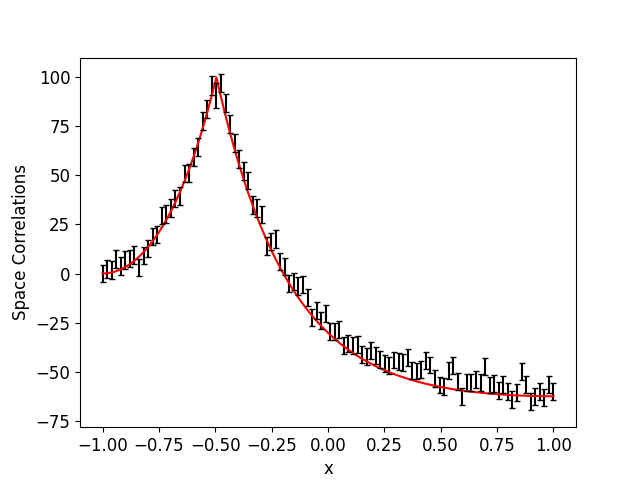}
     \caption{Spatial dependence of the difference between the pair correlation function
     evaluated in $(x, y=-0.5)$ and $(x=-L=-1, y=-0.5)$. The black curve
     is given by a Monte Carlo simulation of the $N$-control scheme, while the red one
     is given by the analytical solution of the anarchic model.
     Here, $\theta = 10^{-3}$. Parameters are taken from
   Table~\ref{tab:params_table}.}
     \label{fig:comparison_g_space_a_mc}
 \end{figure}
 
On a side note, Monte Carlo simulations of the
$N$-control model for $\theta \ll 1$ seem to show that the
relative spatial dependence of the pair correlation function is almost exactly
the same as in the anarchic model for $t_1 = t_2$. This is illustrated in
Fig.~\ref{fig:comparison_g_space_a_mc}, where we plot the shape of the correlation
function for both models, for a fixed value of $y=-0.5$ and relative to its value in
$x=-L=-1$. This remark does not hold true for $\theta \sim 1$.

\begin{figure}[t]
    \centering
    \includegraphics[width=\linewidth]{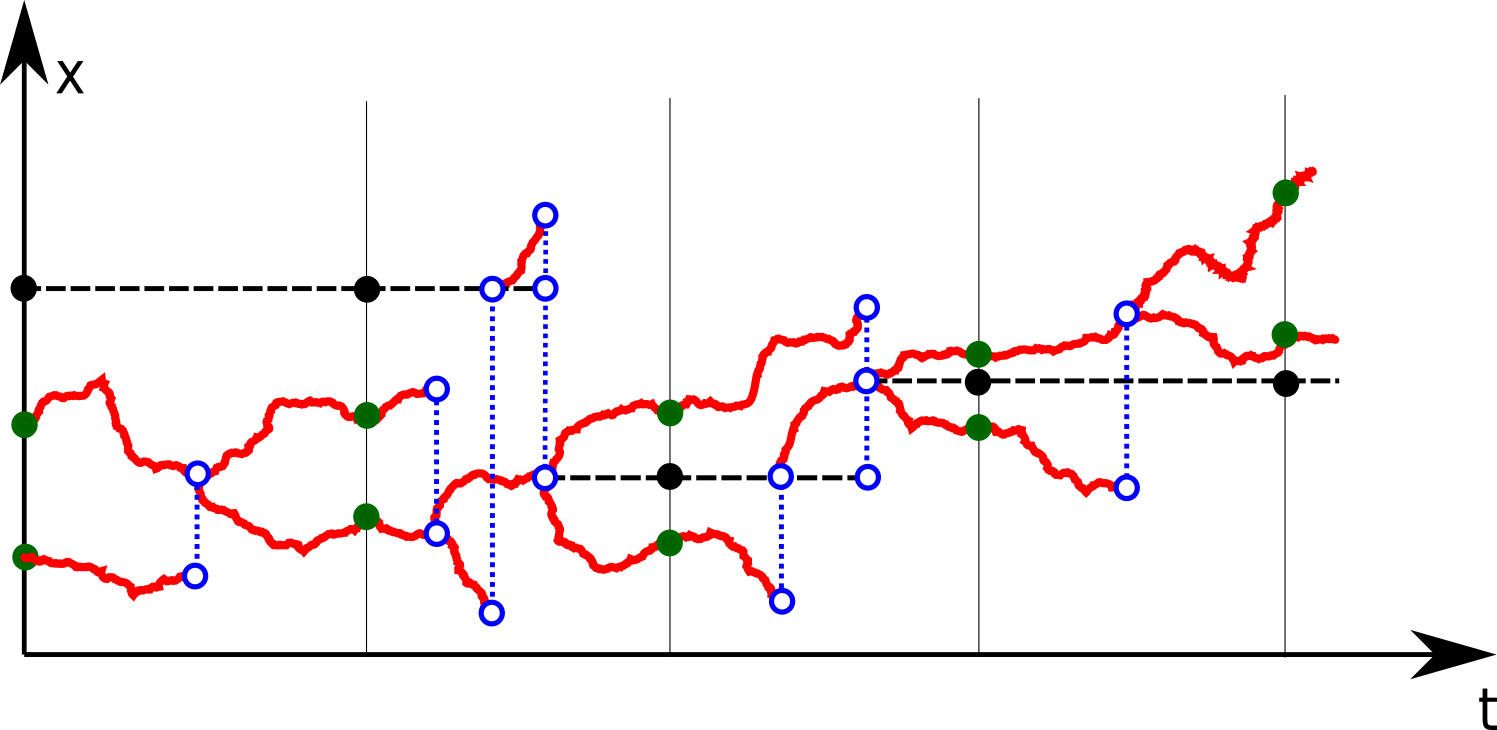}
    \caption{A typical history for a system of 2 neutrons and 1 precursor in
    the $NM$-control model. Neutrons diffuse (in red) until they branch or are
  destroyed by another neutron being produced in the system (these correlated
event are shown in blue dashed lines), and precursor life-time is shown by
black dashed-lines.}
    \label{fig:complete_control_drawing}
\end{figure}

Note that, although the hierarchy of the moment equations is not closed,
the underlying stochastic process can be straightforwardly
simulated using the Monte Carlo method. In the following, we shall make
extensive use of this fact to compare the $N$-control model, which plays the
role of a reference for population control, with other models enforcing
similar constraints on the population. Indeed, the occurrence of
$H^{nn}(x,y,t)$ and
$H^{nm}(x,y,t)$ suggests that the lack of closure originates from the fact that 
the $N$-control model correlates the size of the precursor population
with the local neutron and precursor densities. It is tempting to suggest the use of the approximation
\begin{equation}
H^{n}(x,t) \simeq n(x,t) m(t)
\end{equation}
in order to close Eqs.~\eqref{eq:N_cst_1st}, and of similar approximations
for $H^{nn}(x,y,t)$ and $H^{nm}(x,y,t)$. Doing so in the general case
yields equations with time-dependent coefficients in the moment equations. However,
we do not know how to construct a Monte Carlo game associated to these
approximate moment equations. This effectively prevents us from comparing
the analytical results that we would eventually obtain with the
corresponding simulation results. 

Another way to close the hierarchy of moment equations is to apply
population control to both populations, so that the size of the precursor population
is not a random variable. By doing so, we sacrifice some of the
physical relevance of the $N$-control model, but we gain the ability to
derive analytical solutions while keeping an exact associated Monte Carlo
game. We henceforth propose two alternative models and we discuss under
which conditions they may satisfyingly approximate the
$N$-control model.

\subsection{Control of neutrons and precursors}
\label{sss:NM_cst}

Let us now consider a process in which the total number of neutrons $N$
\textit{and} the total number of precursors $M$ are kept under constraint.
Conservation of $N$ is enforced as previously described. Conservation of $M$ is
enforced as follows: when a neutron undergoes fission and produces a precursor,
another randomly chosen precursor is destroyed. When a precursor emits a delayed
neutron, the precursor is \textit{not} destroyed. In this way, we ensure that $N$ and
$M$ are kept constant at
any time (see Fig.~\ref{fig:complete_control_drawing}).
We call this algorithm the \emph{$NM$-control} scheme.

For the total population statistics, $NM$-control
trivially yields one possible state, given by the initial
condition $(N,M)$, with vanishing variance. It appears however that the
spatial behaviour
is non-trivial. Additionally, direct calculation confirms that the constraint
of fixed $(N,M)$ indeed sidesteps
the closure issues of the $N$-control model. The
equations for the first-order moments are now closed and read:
\begin{subequations}
    \begin{align}
        \begin{split}
            \frac{\partial}{\partial t} n(x,t) &= {\cal D} \nabla^2_x n(x,t) + \lambda \left(m(x,t) - n(x,t)
            \frac{M}{N} \right)
            \label{eq:NM_cst_1st_n}
        \end{split}\\
        \begin{split}
            \frac{\partial}{\partial t} m(x,t) &= \beta  \nu_{d,1} \left( n(x,t) - m(x,t) \frac{N}{M} \right)\text.
            \label{eq:NM_cst_1st_m}
        \end{split}
  \end{align}%
  \label{eq:NM_cst_1st}%
\end{subequations}%
If $N$ and $M$
are in the equilibrium ratio $N/M = \theta = \lambda/(\beta \nu_{d,1})$, then
Eqs.~\eqref{eq:NM_cst_1st} reduce to the usual forward equations for the first
moments of freely evolving populations of neutrons and precursors
\cite{houchmandzadehNeutronFluctuationsImportance2015}. This has two
consequences. First, we can conclude that the mixing time for the $NM$-control
model is the same as for the anarchic model, $\tau_1$. Second,
Eqs.~\eqref{eq:NM_cst_1st} admit uniform and constant asymptotic solutions
\begin{subequations}
\begin{align}
    n(x,t) & \to n_{\infty}(x) = \frac{N}{2L} \\
    m(x,t) & \to m_{\infty}(x) = \frac{M}{2L} \text.
\end{align}%
\label{eq:NM_asympt_1st}%
\end{subequations}%

For the second-order moments we obtain once again a closed system of equations:
\begin{widetext}
\begin{subequations}
    \begin{align}
        \begin{split}
            \frac{\partial}{\partial t} \Tilde{u}(x,y,t) &= {\cal D}\left( \nabla^2_x + \nabla^2_y\right)\Tilde{u}(x,y,t) - \tau_n^{-1}\Tilde{u}(x,y,t) + \lambda C_{N} \left(v(x,y,t) + v(y,x,t)\right)\\
            &\qquad + 2 \beta \delta(x-y) n(x,t)\label{eq:NM_cst_2nd_u}
        \end{split}\\
        \begin{split}
            \frac{\partial}{\partial t} v(x,y,t) &= {\cal D}\nabla^2_x v(x,y,t) - \tau_c^{-1} v(x,y,t) + \beta  \nu_{d,1} C_{N-1} \Tilde{u}(x,y,t) + \lambda \Tilde{w}(x,y,t)
            \label{eq:NM_cst_2nd_v}\\
            & \qquad + \delta (x-y) \left( 2\beta  \nu_{d,1} n(x,t) + \lambda m(x,t)\right)
        \end{split}\\
        \begin{split}
            \frac{\partial}{\partial t} \Tilde{w}(x,y,t) &= \beta  \nu_{d,1} C_{M} (v(x,y,t) + v(y,x,t)) - \tau_p^{-1} \Tilde{w}(x,y,t)\text.
            \label{eq:NM_cst_2nd_w}
        \end{split}
    \end{align}%
    \label{eq:NM_cst_2nd}%
\end{subequations}%
\end{widetext}
with $C_M = (M-1)/M$ and where we defined the time constants
\begin{subequations}%
\begin{align}
    \tau_{n} &= \left(\frac{2 \beta}{N-1} + \frac{2 \lambda M}{N} \right)^{-1} \label{eq:nm_constants_n} \\
    \tau_{c} &= \left(\frac{\beta  \nu_{d,1} N}{M} + \lambda \frac{M}{N} \right)^{-1} \\
    \tau_{p} &= \frac{M}{2 \beta \nu_{d,1} N},
\end{align}%
\label{eq:nm_constants}
\end{subequations}
Equations~\eqref{eq:NM_cst_2nd} are a linear system of equations for
$(\Tilde{u}, v, \Tilde{w})$. Following the heuristic arguments of Zhang et al.\
\cite{zhangDiffusionReproductionProcesses1990}, we should expect the renewal time for
our system to be determined by the dominant time constant of the correlation functions.
In the $NM$-control model, the collective modes of the correlation functions are
associated to the eigenvalues of the matrix of the coefficients of $(\Tilde{u}, v,
\Tilde{w})$ in Eqs.~\eqref{eq:NM_cst_2nd}, which reads
\begin{equation}
  R=\begin{pmatrix}
    -\tau_n^{-1} & 2\lambda C_N & 0 \\
    \beta \nu_{d,1} C_{N-1} & -\tau_c^{-1} & \lambda \\
    0 & 2 \beta\nu_{d,1} C_M & -\tau_p^{-1}
  \end{pmatrix}
  \text.
\end{equation}
Assuming that $N$ and $M$ are chosen in the equilibrium ratio $N=\theta M$ and
that $N$ is large, we can extract the scaling behaviour of the
eigenvalues of $R$ with respect to $\theta$, for small $\theta$. The
eigenvalues $\{r_1,r_2,r_3\}$ read
\begin{align}
  r_1 &{}\simeq-2\beta\nu_{d,1}\\
  r_2 &{}\simeq-\beta\nu_{d,1}\\
  r_3 &{}\simeq-\frac{2\beta\theta^2(1+3\nu_{d,1})}{N(1+\theta)^2}
  \text,
\end{align}
with $r_1<r_2<r_3$. The time constant associated to the dominant eigenvalue,
$r_3$, is the renewal time for the $NM$-control model, and reads
\begin{equation}
  \tau_R^{NM} \simeq \frac{N(1+\theta)^2}{2\beta\theta^2(1+3\nu_{d,1})}
  \text.
\end{equation}
This expression is very similar to the extinction time for the anarchic scheme,
Eq.~\eqref{eq:extinction_time_anarchic}. However, the renewal time
$\tau_R^{NM}$ is not related to the extinction time for the system, because no
critical catastrophe is possible in the $NM$-control model.

As we did for the anarchic scheme, we can then define a dimensionless
clustering parameter $\eta^{NM} = \tau_1 / \tau_R^{NM}$. However, given that
the mixing time of the $NM$-control scheme is the same as the mixing time of
the anarchic scheme, and that the renewal time $\tau_R^{NM}$ is very
similar to the extinction time of the anarchic scheme, the clustering parameter
also results in a $1/\theta$ scaling for small $\theta$, and the remarks made
about the anarchic scheme apply verbatim to the $NM$-control scheme.

The full solution for Eqs.~\eqref{eq:NM_cst_2nd} is
cumbersome. However, we can obtain asymptotic solutions in the form of Fourier series.
Let $\Tilde{u}_{\infty}(x,y)$, $v_{\infty}(x,y)$ and $\Tilde{w}_{\infty}(x,y)$ be the
asymptotic shapes of the correlation functions for long times:
\begin{align*}
    \Tilde{u}_{\infty}(x,y) &{}= \lim_{t\to\infty} \Tilde{u}(x,y,t)\\
    v_{\infty}(x,y) &{}= \lim_{t\to\infty} v(x,y,t)\\
    \Tilde{w}_{\infty}(x,y) &{}= \lim_{t\to\infty} \Tilde{w}(x,y,t)
    \text.
\end{align*}
We use the following Fourier decomposition:
\begin{equation}
  f(x,y) = \sum_{k_x,k_y=-\infty}^{+\infty}
  f_{k_x,k_y}\exp\left(i \kappa_{k_x}\,x\right)\exp\left(i \kappa_{k_y}\,y\right)
  \text.
  \label{eq:fourier_coeffs}
\end{equation}
Here $\kappa_k= k \pi /( 2 L)$ are the characteristic wave numbers for reflection
boundary conditions and $f(x,y)$ stands for any of
$\Tilde{u}_{\infty}(x,y)$, $v_{\infty}(x,y)$ or $\Tilde{w}_{\infty}(x,y)$. The
coefficients satisfy the relation
\begin{multline}
f_{k_x,k_y} =\\
\frac{1}{2\pi}
\iint \exp\left(-i \kappa_{k_x} \, x\right)
\exp\left(-i \kappa_{k_y} \, y\right) f(x,y)
\,dx\,dy
\text.
\label{eq:fourier_coeff_def}
\end{multline}
The linear system of equations solved by the Fourier coefficients can be obtained from
Eqs.~\eqref{eq:NM_cst_2nd} by setting the time derivative terms to zero, multiplying by
$\exp(-i \kappa_{k_x})\exp(-i \kappa_{k_y}y)/(2\pi)$, integrating over
$dx\,dy$ and using Eq.~\eqref{eq:fourier_coeff_def} and Eqs.~\eqref{eq:NM_cst_1st}. The
analytical expressions of the coefficients are given in Eqs.~\eqref{eq:NM_sol}.

\begin{figure}
  \centering
  \includegraphics[width=\linewidth]{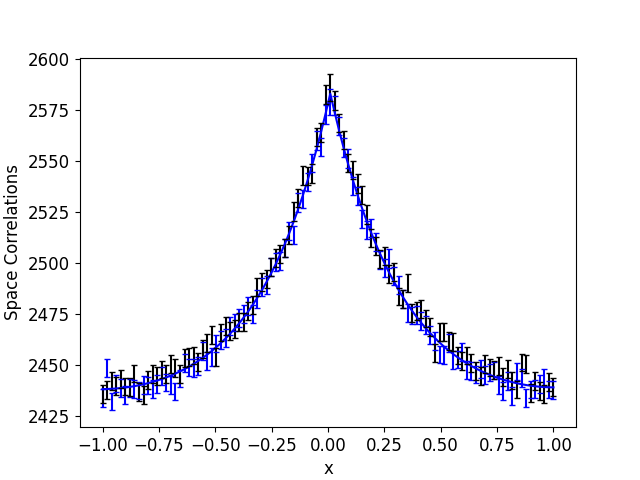}
  \caption{Asymptotic pair correlation function for $\theta = 10^{-3}$ and $y=0$. The parameters are taken from Table~\ref{tab:params_table}. Black line: analytical solution of the $NM$-control
  scheme, Eqs.~\eqref{eq:NM_cst_2nd}.
  Black with error bars: Monte Carlo simulations for the
  $N$-control scheme. Blue with error bar: Monte Carlo simulations of the $NM$-control
  scheme, both with $10^6$ replicas.}
  \label{fig:complete_space_th0001}
\end{figure}

\begin{figure}
  \centering
  \includegraphics[width=\linewidth]{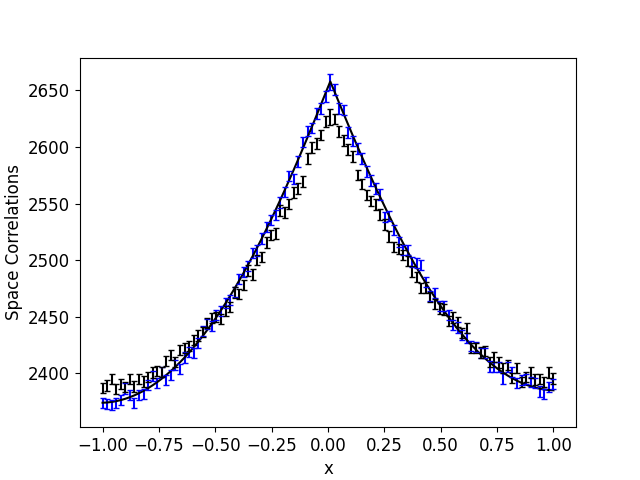}
  \caption{Same color code as Fig.~\ref{fig:complete_space_th0001} but the parameters are taken from Table~\ref{tab:params_table} for $\theta = 1$.}
  \label{fig:complete_space_th1}
\end{figure}

The asymptotic two-point neutron-neutron correlation function for
the $NM$-control model was numerically computed by truncating the Fourier series after
1000 terms.
It is compared with Monte Carlo simulations for the $N$-control
and the $NM$-control scheme, for two sets of physical parameters, in
Figs.~\ref{fig:complete_space_th0001} and \ref{fig:complete_space_th1}.

For $\theta = 0.001$ (Fig.~\ref{fig:complete_space_th0001}),
the $NM$-control scheme provides a good approximation of the $N$-control scheme, and all
the curves representing neutron correlations are very similar.
This means that, when $\theta \ll 1$, which is
the case in nuclear
reactors, a system where neutrons and precursors are both kept under
control has similar correlation functions as a system where population control
acts on neutrons alone. On the other hand, when $\theta = 1$
(Fig.~\ref{fig:complete_space_th1}), the
$N$-control and $NM$-control models yield similar, but different correlation functions.
Indeed, we observe that the $NM$-control scheme results in stronger spatial short-range
correlations compared to the $N$-control scheme, while long-range correlations are
weaker in the $NM$-control scheme. This can be easily explained: in
the $N$-control scheme, when a precursor decays, it is destroyed
and replaced by a neutron that immediately starts diffusing.
This mechanism can be seen as a kind of delayed diffusion; when a
neutron induces a fission event and produces a precursor, it effectively suspends
diffusion for a time of the order of $1 / \lambda$. On the other hand, in the
$NM$-control model, a precursor does not die when it decays, and thus plays the role
of a source at a fixed position until it is replaced by another precursor. The precursor
density can then undergo significant local fluctuations favouring short-range
correlations by inducing overproduction of neutrons
in a small volume, at the expense of long-range correlations.

Some insight about the similarity between the $N$-control and $NM$-control schemes for
$\theta \ll 1$ can be gained by
comparing Eqs.~\eqref{eq:N_cst_2nd} and Eqs.~\eqref{eq:NM_cst_2nd}.
Indeed, remember that $\theta \ll 1$ means that the number of precursors is much larger
than the number of neutrons. If $m(t)$ is sufficiently large, which is the
case here, its
fluctuations will be negligible compared to its mean value.
This allows us to
consider $m(t)$ as a deterministic observable. If we then assume that
\begin{equation*}
    \mathbb{E} [n_i(t) n_j(t) m(t)] \simeq \mathbb{E} [n_i(t) n_j(t)] \mathbb{E}[m(t)]
    \text,
\end{equation*}
then Eq.~\eqref{eq:N_cst_2nd_u} reduces to Eq.~\eqref{eq:NM_cst_2nd_u}. Thus, in this
limit, neutron-neutron correlations follow the same dynamics in both schemes.
It is worth remarking however that, under the same approximation, the equations for $v$
and $\Tilde{w}$ in the $N$-control model do not reduce to the corresponding equations in
the $NM$-control model.
 
\subsection{Immigration model}
\label{sss:imm}

\begin{figure}
    \centering
    \includegraphics[width=\linewidth]{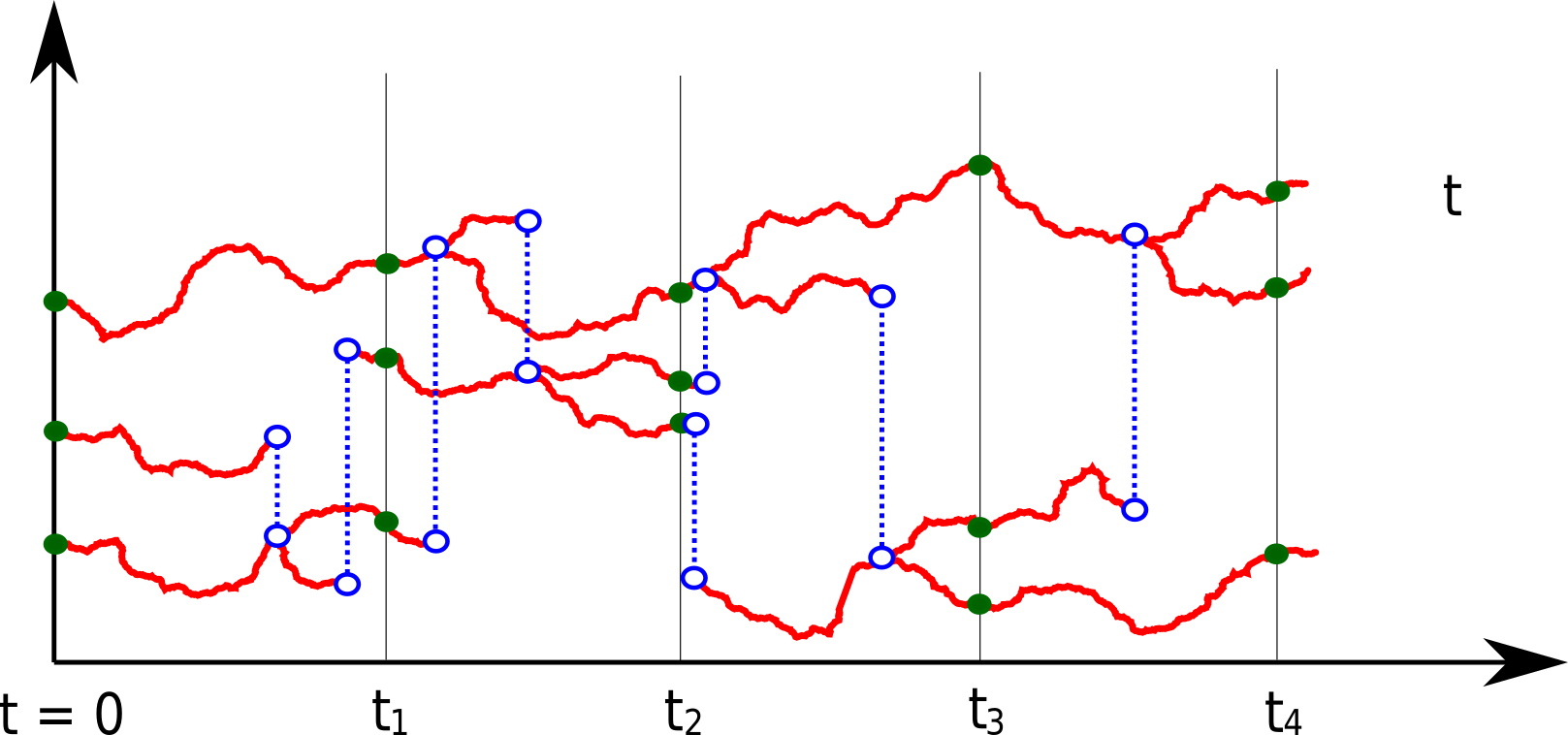}
    \caption{A typical history for a system with 3 neutrons at $t = 0$, in the immigration model. Neutrons diffuse (in red) until they branch or are killed by another neutron branching somewhere else (blue lines show these correlated events). They can also be produced by the source, thus killing another randomly chosen neutron.}
    \label{fig:immigration_drawing}
\end{figure}

We have observed that the precursor population only exhibits weak fluctuations in the
$\theta \ll 1$ limit. This suggests an alternative model where precursors are
modeled as a fixed external source, with the aim of obtaining a time-dependent
solution that, in this limit, successfully approximates the $N$-control model. Thus, we
model the system as a collection of neutrons under population control, with an external,
time-independent source term modelling precursor decay. Formally, this model is
equivalent to an {\em immigration model} with a time-independent, Poissonian immigration source
describing the asymptotic precursor decay density, given by
\begin{equation}
    {\cal Q}_I(x) = \lambda m_\infty(x) = \frac{\lambda }{\theta}n_\infty(x)\text,
    \label{eq:immigration_source}
\end{equation}
where $\theta$ is the parameter of the
equivalent $NM$-control scheme. Population control is enforced by requiring that each
time a new neutron enters the system, whether by fission or from the external source,
we destroy another randomly chosen neutron, as illustrated in
Fig.~\ref{fig:immigration_drawing}. 

Following the usual derivation method,
we obtain the equation for the average neutron density, namely
\begin{equation}
    \frac{\partial}{\partial t} n(x,t) = {\cal D}\nabla^2 n(x,t)
    - \frac{\lambda}{\theta} n(x,t) + \mathcal{Q}_I(x)
    \text.
    \label{eq:imm_1st}
\end{equation}
Taking $\mathcal{Q}_I(x)$ as given by Eq.~\eqref{eq:immigration_source}, the last two
terms asymptotically cancel out and
the dynamics of the average is driven only by diffusion, similarly to what happens in
the anarchic model when the system is critical.

The equation for the pair correlation function reads 
\begin{multline}
    \frac{\partial}{\partial t} \Tilde{u}(x,y,t) = {\cal D}\left(\nabla^2_x + \nabla^2_y 
    \right)\Tilde{u}(x,y,t) - \tau_{n}^{-1} \Tilde{u}(x,y,t)\\
    + C_{N} \left(n(x,t)\mathcal{Q}_I(y) + n(y,t)\mathcal{Q}_I(x) \right)\\
    + 2\beta \delta(x-y) n(x,t),
   \label{eq:imm_2nd}
\end{multline}
where we recognize $\tau_{n}$ from Eq.~\eqref{eq:nm_constants_n}. Thus, in this model,
neutron pair correlations relax with a time constant governed by $\tau_n$.

Equations~\eqref{eq:imm_1st}
and \eqref{eq:imm_2nd} have the same form as Eqs.~\eqref{eq:NM_cst_1st_n} and
\eqref{eq:NM_cst_2nd_u}, provided that one replaces $\mathcal{Q}_I(x)$ by its
expression, Eq.~\eqref{eq:immigration_source}. These
equations can be solved analytically. In particular, taking a uniform
initial condition for $n(x,t)$, we obtain
\begin{equation}
  n(x,t) = \frac{N}{2 L}
  \label{eq:immigration_n}
\end{equation}
for the average neutron density, and 
\begin{multline}
    \Tilde{u}(x,y,t) = \frac{N (N-1)}{4 L^2} \exp\left(-\frac{t}{\tau_{n}}\right)\\
    + \frac{2 \tau_{n} \lambda M (N-1)}{4 L^2}
    \left[1-\exp\left(-\frac{t}{\tau_{n}}\right)\right] \\
    + \frac{2\beta N}{2 L}\int dx' \int_0^t dt' {\cal G'}(x,t|x',t') {\cal G'}(y,t|x',t')\\
    \exp\left(-\frac{t-t'}{\tau_{n}}\right)
    \label{sol:imm_2nd}
\end{multline}
for the pair correlation function, where ${\cal G'}$ is the Green's function associated to
Eq.~\eqref{eq:imm_1st}. As a side note, taking the $\lambda \to 0$ limit
in this equation
we recover the pair correlation function for a system of prompt neutrons under
population control \cite{meyerClusteringIndependentlyDiffusing1996,
zoiaClusteringBranchingBrownian2014, zoiaNeutronClusteringSpatial2017}.

\begin{figure}
     \centering
     \includegraphics[width=\linewidth]{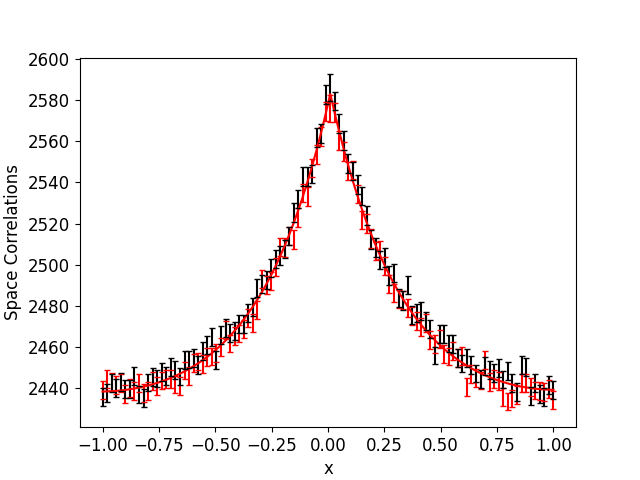}
     \caption{Asymptotic pair correlation function for $\theta = 0.001$ and $y=0$.
     Parameters are the same as in Fig.~\ref{fig:complete_space_th0001}. Red curve:
     analytical solution
     of the immigration model. Red error bars: Monte Carlo simulations of the
     immigration model. For reference, the results of
     the $N$-control scheme are plotted in black.} 
     \label{fig:imm_space_th0001}
 \end{figure}

\begin{figure}
  \vspace{0.80cm}
  \centering
  \includegraphics[width=\linewidth]{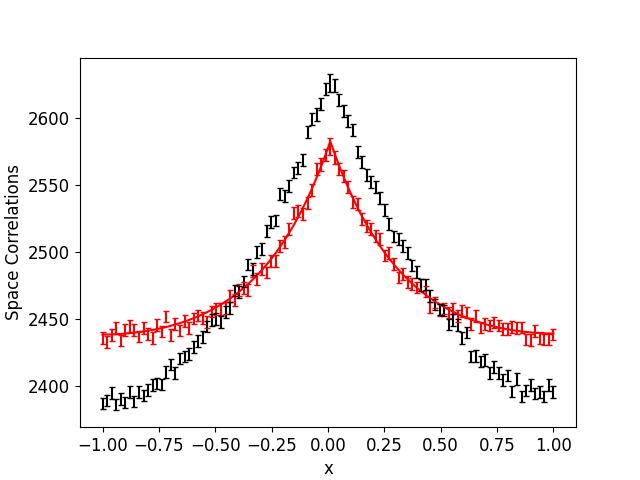}
  \caption{Same color code than Fig.~\ref{fig:imm_space_th0001}, and parameters corresponding to $\theta = 1$ in Table~\ref{tab:params_table}.} 
  \label{fig:imm_space_th1}
\end{figure}

Comparison with Monte Carlo simulations (see Fig.~\ref{fig:imm_space_th0001})
shows that, for systems where $\theta\ll 1$, the immigration model closely
matches the $N$-control model; the remarks made about
Fig.~\ref{fig:comparison_g_space_a_mc} also apply to the immigration model.
Figure~\ref{fig:imm_space_th1} provides the same
comparison for a larger value of $\theta$. Comparing Fig.~\ref{fig:imm_space_th1} and
Fig.~\ref{fig:complete_space_th1}, it is clear that the $N$-control model is better
approximated by the $NM$-control model than by the immigration model. This is
unsurprising, in view of the crudeness of the treatment of neutron-precursor
correlations in the immigration model.

To summarize the results of this section, we have modeled feedback as an idealized
mechanism enforcing strict conservation of the total number of neutrons. We showed that
doing so raises a problem of non-closure of the
associated hierarchy of moment equations. Enforcing population control on neutrons and
precursors solves the closure problem, but results in a less physically realistic model,
especially if the condition $\theta \ll 1$ does not hold. We also showed that it is
possible to model precursors as a fixed source term related to the asymptotic precursor
density, which also results in a satisfactory approximation of the neutron-precursor
model for $\theta \ll 1$.

\subsection{Mean-squared pair distance functions}

We now discuss the mean-squared pair distance for our population control models, and we
compare it to the case of an anarchic population.

For the $N$-control scheme, we do not have access to an analytical expression for the
mean-squared pair distance, because the moment equations, Eqs.~\eqref{eq:N_cst_2nd},
are not closed. However, we do have an asymptotic correlation function for the
$NM$-control scheme, given by Eqs.~\eqref{eq:NM_sol}, from which we can deduce an
asymptotic mean-squared pair distance. For reflection boundary conditions, it is given
by
\begin{equation}
    \langle r^2_{NM} \rangle = \frac{8 L^4 \Tilde{u}_{0,0}}{3 N^2}
    - \left(\frac{16 L^2}{\pi^2 N}\right)^2 \, \sum_{k=1}^{+\infty} \frac{\Tilde{u}_{k,k}}{k^4}
    \text,
    \label{eq:asymp_NM_r}
\end{equation}
where $\Tilde{u}_{k,k}$ is the Fourier coefficient
for the neutron-neutron correlations asymptotic solution of Eq.~\eqref{eq:NM_cst_2nd_u}.
On physical grounds, we expect the asymptotic pair distance in the
$NM$-control scheme to be always smaller than the uncorrelated value,
Eq.~\eqref{eq:uncorrelated_pair_dist}.

\begin{figure}
    \centering
    \includegraphics[width=\linewidth]{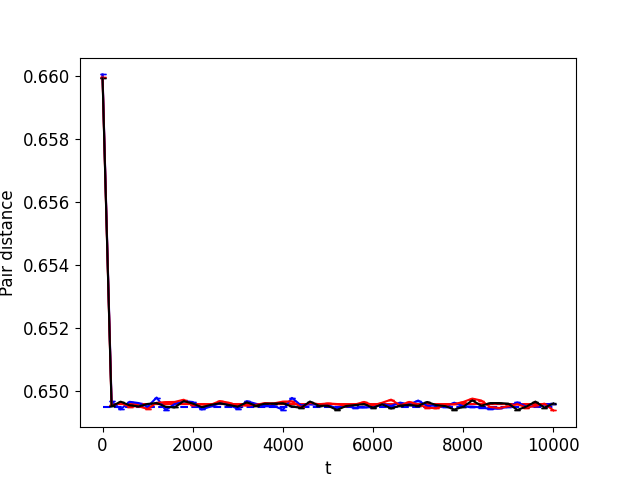}
    \caption{Mean-squared pair distance for the same parameters as
    Fig.~\ref{fig:complete_space_th0001} ($\theta = 0.001$). Blue dashed line:
  $NM$-control model analytical asymptotic value; blue with error bar: Monte
Carlo result; red dashed line: immigration model analytical solution; red with
error bar: Monte Carlo result; black: $N$-control model Monte Carlo result. All
simulation results are obtained with $10^6$ replicas.}
    \label{fig:pair_distance_control_th0001}
\end{figure}

\begin{figure}
    \includegraphics[width=\linewidth]{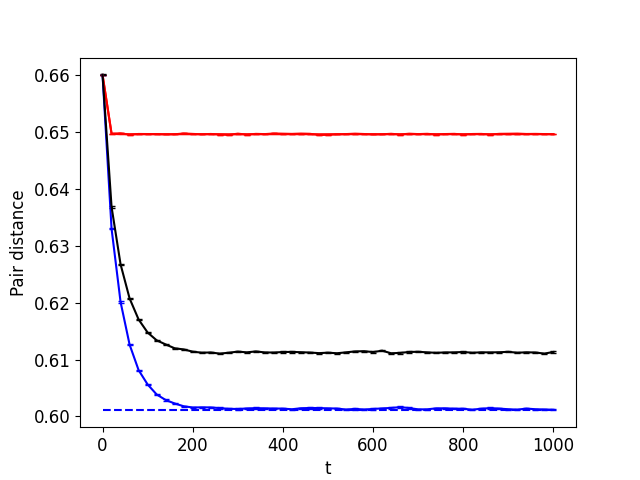}
    \caption{Mean-squared pair distance with same parameters as in Fig.~\ref{fig:complete_space_th1} ($\theta = 1$), and color code of Fig.~\ref{fig:pair_distance_control_th0001}.}
    \label{fig:pair_distance_control_th1}
\end{figure}

As for the immigration model, an analytical expression for the mean-squared pair
distance can be obtained by applying the definition of the mean-squared pair distance
function, Eq.~\eqref{eq:def_msd}, to the solution, Eq.~\eqref{sol:imm_2nd}. For
reflection boundary conditions, straightforward calculations hence yield
\begin{equation}
    \langle r^2_I \rangle(t) = C_N\frac{2 L^2}{3} - \frac{128 L^2
    \beta}{N}\sum_{k=1}^{+\infty} \frac{1-e^{-\left(\tau_{n}^{-1} -
    2\alpha_k\right)t}}{(k \pi)^4 \left( \tau_{n }^{-1} - 2\alpha_k\right)}
    \label{eq:r2_immigration}
\end{equation}
The sum can be computed analytically in the asymptotic limit and reads
\begin{multline}
    \langle r^2_I \rangle_\infty = C_N\frac{2 L^2}{3} + \frac{16 \beta
    \tau_n}{N}\times\\
    \left(\left(\frac{D \tau_n}{L }\right)^2
     - \frac{2 D \tau_n}{3 } - \frac{4 L^2}{45} - \frac{\sqrt{2} (D\tau_n)^\frac{3}{2}}{L} \cot{\sqrt{\frac{2 \tau_n}{D}} L} \right)
     \label{eq:r2_immigration_asympt}
\end{multline}
If we take the mean-squared pair distance as an indicator of clustering,
Fig.~\ref{fig:pair_distance_control_th0001}
clearly illustrates that the three models have very similar clustering behaviour for
$\theta \ll 1$, even for very long times. For $\theta \sim 1$, on the other hand, the
models behave differently, as shown in Fig.~\ref{fig:pair_distance_control_th1}. In
particular, we note that the immigration model yields the largest mean-squared pair
distance of all the considered control schemes. This is easily explained by the fact that
the ``decay source'' of the immigration model is assumed to be uniformly distributed in
space and completely uncorrelated with the neutrons, an assumption that is bound to
reduce clustering. The $NM$-control scheme yields a smaller mean-squared
pair distance than the $N$-control scheme, indicating that clustering is more
prominent in the former. This behavior can be explained by observing that a
precursor may in fact produce several neutrons at the same position before being
destroyed, which tends to increase clustering. It is interesting to note that
for all control models 
the squared pair distance asymptotically converges to
$\langle r^2 \rangle_{\infty} < \langle r^2\rangle_{\text{iid}}$,
because the critical catastrophe is avoided, contrary to what happens in the anarchic case.
However, Eqs.~\eqref{eq:asymp_NM_r} and \eqref{eq:r2_immigration_asympt} show
that $\langle r^2 \rangle_{\infty}$ tends to the uncorrelated value as $N$
tends to infinity.

\section{Conclusions}
\label{sec:conclusion}

We considered a simplified one-dimensional model of a nuclear reactor
including neutrons and precursors.
We derived moment equations characterizing the
spatial correlations of our system.
Our main tool was the two-point, two-time neutron-neutron
second-order moment (hereafter named pair correlation function), which characterizes the
joint observation of neutrons at $(x_1,t_1)$ and at $(x_2,t_2)$. By
using the P\'al and Bell backward formalism, we derived the pair correlation function for
the anarchic model, where all the populations are left free to evolve. We identified two
typical time scales, namely the time $\tau_2$ necessary for the prompt correlation
dynamics to relax, and the time $\tau_1$ necessary for correlations to achieve their
asymptotic spatial shape. The latter generalizes the concept of mixing time
to a system with delayed neutrons; however, for typical values of the physical
parameters, $\tau_1$ is actually much longer than the mixing time of a critical system
without delayed neutrons. The system is furthermore characterized by its extinction
time $\tau_E$, which represents the typical time required to observe catastrophic
fluctuations of the total population size. Similarly to the mixing time, the extinction time
for a system with delayed neutrons is much longer than the extinction time for a
critical system without delayed neutrons, for typical values of the physical parameters.

In an attempt to take into account feedback effects, we introduced several
idealized population control mechanisms. These models break the statistical
independence of the neutron fission chains, which forced us to abandon the
P\'al and Bell backward formalism in favor of the forward formalism. We derived
the master equations for three different population control models:
\begin{itemize}
    \item in the $N$-control model,
    the neutron population is kept constant, but no constraint
    is applied to the precursor population. This model yields a non-closed hierarchy of
    moment equations;
    \item in the $NM$-control model, both the neutron and precursor populations
    are controlled. This model yields a closed hierarchy of moment equations.
    We identified the dominant time scales of the model and presented
    asymptotic solutions for the first two moment orders.
    \item in the immigration model, precursors are modeled as a uniform Poisson neutron
    source whose intensity is equal to the asymptotic precursor decay rate at equilibrium.
\end{itemize}
For each model, we obtained two-point, one-time pair correlation functions by Monte Carlo simulations,
and by analytical means when possible. Comparisons
between analytical solutions and Monte Carlo simulations showed that
all the models yield
very close results when $\theta \ll 1$, which is the main regime of
interest for nuclear reactors. We also showed that the introduction of population
control prevents the occurrence of the critical catastrophe and thus quenches clustering.

As a prospect for future work, it is of course interesting to consider the possibility
of lifting the simplifying assumptions of our current model, e.g.\ by introducing
energy- and space-dependent neutron interaction rates, and by replacing the diffusion
approximation
(Brownian motion) with transport (jump processes with random redistribution of directions).
These refinements, however, would probably bring the models out of reach of an
analytical
treatment. More interestingly, population control acts as a global
mechanism in our model, whereas in reality feedback effects are local. This suggests the
study of spatially-dependent versions of the population control mechanisms considered here.
Finally, the extension of the
present work to non-critical regimes is relevant to reactors in power excursions (e.g.,
during start-up and shut-down), and is also particularly interesting for ecological,
epidemiological and biological models.

\appendix

\begin{widetext}

\section{Backward formalism for the anarchic model}
\label{appendix:backward_formalism}

Consider the branching Brownian motion whose stochastic rules are given in
Sec.~\ref{sec:free_pop} and assume that no constraint is enforced on the total population.
Let $V_1$ and $V_2$ be two finite-size non-overlapping detectors, and let ${\cal
P}(n_1,t_1,n_2,t_2|x_0,t_0)$ be the probability to find $n_1$ particles at time $t_1$ in
detector $V_1$ and $n_2$ particles at time $t_2$ in detector $V_2$, for a single particle
starting in $x_0$ at time $t_0$, with $t_2 > t_1$; the results for $t_2<t_1$
are recovered by arguments of symmetry. P\'al and Bell have independently derived
a backward formalism yielding the (adjoint) master equation for ${\cal P}$. The
manipulation
of the resulting equation is rather cumbersome, and in most cases neutron noise can be
characterized using the lowest-order
moments of $n_1$ and $n_2$. We will be in particular interested in the average number of
neutrons in detector $V_1$ at time $t_1$, namely,
\begin{equation}
    \mathbb{E}[n_1](t_1|x_0,t_0)=
        \sum\limits_{n_1, n_2} n_1 \mathcal{P}(n_1,t_1,n_2,t_2|x_0,t_0)
\end{equation}
and in the two-point correlation function between particles detected in $V_1$ at time $t_1$ and particles detected in $V_2$ at time $t_2$, namely,
\begin{equation}
    \mathbb{E}[n_1 n_2](t_1, t_2|x_0,t_0)=
        \sum\limits_{n_1, n_2} n_1\,n_2\,\mathcal{P}(n_1,t_1,n_2,t_2|x_0,t_0).
\end{equation}
By virtue of these considerations, it is therefore more convenient to introduce the probability-generating function associated to ${\cal P}$, namely,
\begin{equation}
W(z_1,z_2,x_0,t_0) = \sum_{n_1} \sum_{n_2} z_1^{n_1} z_2^{n_2} P(n_1,n_2, t_1, t_2|x_0,t_0),
\end{equation}
where $z_i$ is the conjugate variable of $n_i$, $i \in \{ 1,2 \}$, and hence derive the equations for the (falling factorial) moments, which follow from
\begin{equation}
\mathbb{E}[n_1(n_1-1)\ldots(n_1-k+1)n_2(n_2-1)\ldots(n_2-l+1)](t_1,t_2|x_0,t_0) = \frac{\partial^{k+l}}{\partial z_1^k \partial z_2^l} W(z_1,z_2,x_0,t_0 )\vert_{z_1=1,z_2=1},
\label{moment_der}
\end{equation}
with $k,l \ge 0$ and the normalization $W(1 , 1,x_0,t_0 ) = 1$. It can be shown that $W(z_1,z_2,x_0,t_0)$ satisfies the non-linear backward evolution equation
\begin{equation}
-\frac{\partial}{\partial t_0} W = {\cal D} \nabla^2_{x_0} W - (\gamma + \beta) W + \beta G_p(W) G_d(W_d) + \gamma,
\label{pal_bell_W2}
\end{equation}
where $G_p(z) = \sum_k z^k p_k$ is the probability-generating function associated to the
prompt fission probability distribution $p_k$ and $G_d(z) = \sum_k z^k q_k$ is the
probability-generating function for the precursor probability distribution $q_k$. The quantity $W_d$ in Eq.~\eqref{pal_bell_W2} is the
probability-generating function associated to the creation of a precursor at $(x_0,t_0)$,
namely,
\begin{align}
&W_d(z_1,z_2|x_0,t_0)  = \lambda \int^{\infty}_{t_0} e^{-\lambda(t'-t_0)} W(z_1,z_2|x_0,t') dt'
\text.
\label{def_W_delayed}
\end{align}
Taking the derivative of Eq.~\eqref{pal_bell_W2} once with respect to $z_1$ and using
the definition in Eq.~\eqref{moment_der}, we obtain the evolution equation for the
average neutron number in $V_1$, for a single starting particle in $(x_0,t_0)$:
\begin{equation}
{\cal L}^\dag \mathbb{E}[n_1](t_1|x_0,t_0) = 0,
\label{eq:pal_bell_ave_appendix}
\end{equation}
where ${\cal L}^\dag$ is the backward linear operator defined in Eq.~\eqref{adj_generator_def}. Equation~\eqref{eq:pal_bell_ave_appendix} (which coincides with Eq.~\eqref{eq:pal_bell_ave} in the main text) is to be solved with the final condition 
\begin{equation}
    \mathbb{E}[n_1](t_1|x_0,t_0=t_1) = \left\{
    \begin{array}{ll}
        1 \enspace & \text{if} \enspace x_0 \in V_1 \\
        0 \enspace & \text{otherwise}\text.
    \end{array}
    \right.
\end{equation}
Letting ${\cal G}(x,t,x_0,t_0)$ be the Green's function associated to the
operator ${\cal L}^\dag$, the solution to
Eq.~\eqref{eq:pal_bell_ave_appendix} reads
\begin{equation}
\mathbb{E}[n_1](t_1|x_0,t_0) = \int dx' \chi_{V_1}(x') {\cal G}(x',t_1,x_0,t_0),
\label{eq:pal_bell_ave_appendix_solution}
\end{equation}
where $\chi_{V}(x)$ is the characteristic function of a domain $V$.

Taking the mixed derivative of Eq.~\eqref{pal_bell_W2} with respect to $z_1$
and $z_2$, and using again the definition in Eq.~\eqref{moment_der}, we
obtain the evolution equation for the two-point correlation function,
for a single
starting particle in $(x_0,t_0)$:
\begin{align}
&{\cal L}^\dag \mathbb{E}[n_1 n_2](t_1,t_2 | x_0, t_0)= -f_\text{corr}(t_1,t_2 | x_0, t_0),
\label{eq:pal_bell_cov_appendix}
\end{align}
where we have defined
\begin{multline}
    f_\text{corr}(t_1,t_2 | x_0, t_0) = \beta \nu_{p,2} \mathbb{E}[n_1]\mathbb{E}[n_2]  +\beta \nu_{p,1} \nu_{d,1} \mathbb{E}[n_1] \lambda \int^{t_2}_{t_0} e^{-\lambda(t'-t_0)} \mathbb{E}[n_2](t') dt'\\
    + \beta\nu_{p,1} \nu_{d,1} \mathbb{E}[n_2] \lambda \int^{t_1}_{t_0} e^{-\lambda(t'-t_0)} \mathbb{E}[n_1](t') dt'
    + \beta\nu_{d,2}  \lambda^2 \int^{t_1}_{t_0} e^{-\lambda(t'-t_0)} \mathbb{E}[n_1](t') dt' \int^{t_2}_{t_0} e^{-\lambda(t'-t_0)} \mathbb{E}[n_2](t') dt' 
\end{multline}
with the shorthand notation $\mathbb{E}[n_i] = \mathbb{E}[n_i](t_1|x_0,t_0)$, $i \in \{ 1,2 \}$. Equation~\eqref{eq:pal_bell_cov_appendix} is to be solved with the final condition
\begin{equation}
    \mathbb{E}[n_1 n_2](t_0=t_1) = \left\{
    \begin{array}{ll}
         \mathbb{E}[n_2](t_2|x_0,t_0=t_1) \enspace & \text{if} \enspace x_0 \in V_1  \\
         0 \qquad & \text{otherwise}\text.
    \end{array}
    \right.
\end{equation}
Using the Green's function associated to the operator ${\cal L}^\dag$, the solution to Eq.~\eqref{eq:pal_bell_cov_appendix} can be expressed as
\begin{multline}
    \mathbb{E}[n_1 n_2](t_1,t_2 | x_0, t_0) = \\
    \int dx' \chi_{V_1}(x') \mathbb{E}[n_2](t_2|x',t_1)  {\cal G}(x',t_1,x_0,t_0)
    + \int_{t_0}^{t_1} dt' \int dx' f_\text{corr}(t_1,t_2 | x', t')  {\cal G}(x',t',x_0,t_0).
    \label{mimj}
\end{multline}
Finally, the case of a collection of ${\cal N}$ independent particles whose initial coordinates are distributed according to a given source ${\cal Q}(x_0, t_0)$ can be dealt with by observing that the resulting probability-generating function $W_{\cal Q}(z_1,z_2)$ is related to the one-particle probability-generating function $W(z_1,z_2,,x_0,t_0)$ by
\begin{equation}
W_{{\cal Q}_{\cal N}}(z_1,z_2) = \left[ \int^{t_1}_{-\infty} dt_0\int dx_0 \,
W(z_1,z_2,x_0,t_0) {\cal Q}(x_0, t_0) \right]^{\cal N}.
\label{pal_bell_W_source}
\end{equation}
Then, by taking the derivative of Eq.~\eqref{pal_bell_W_source} once with respect to $z_1$ we obtain the average number of particles in $V_1$ for a collection of ${\cal N}$ source particles:
\begin{equation}
\mathbb{E}[n_1](t_1|{\cal Q}_{\cal N}) = {\cal N} \int^{t_1}_{-\infty}dt_0 \int d x_0 \mathbb{E}[n_1](t_1 | x_0, t_0)  {\cal Q}(x_0,t_0) .
\label{eq:pal_bell_ave_appendix_solution_source}
\end{equation}
Similarly, by taking the mixed derivative of Eq.~\eqref{pal_bell_W_source} with respect to $z_1$ and $z_2$ we obtain the pair correlation between detector $V_1$ and $V_2$ for a collection of ${\cal N}$ source particles:
\begin{align}
&\mathbb{E}[n_1 n_2](t_1,t_2 |{\cal Q}_{\cal N}) = {\cal N}\left({\cal N}-1 \right) \mathbb{E}[n_1](t_1|{\cal Q}_{\cal N}) \mathbb{E}[n_2](t_2|{\cal Q}_{\cal N}) \nonumber\\
&\qquad+ {\cal N} \int^{t_1}_{-\infty}dt_0 \int d x_0 \mathbb{E}[n_1 n_2](t_1,t_2 | x_0, t_0)  {\cal Q}(x_0,t_0) .
\label{eq:pal_bell_cov_appendix_solution_source}
\end{align}
The continuous version of Eqs.~\eqref{eq:pal_bell_ave_appendix_solution_source} and~\eqref{eq:pal_bell_cov_appendix_solution_source} can be obtained by centering the detector regions $V_1$ and $V_2$ around $x_1$ and $x_2$, respectively, defining
\begin{gather*}
    n(x_1,t_1) = \lim_{V_1\to 0} \frac{\mathbb{E}[n_1](t_1|{\cal Q}_{\cal N})}{V_1}\\
    u(x_1,t_1,x_2,t_2) = \lim_{V_1, V_2\to 0} \frac{\mathbb{E}[n_1 n_2](t_1,t_2|{\cal Q}_{\cal N})}{V_1 V_2}
\end{gather*}
and letting the detector sizes shrink to zero, from which we recover Eqs.~\eqref{eq_ave_density_source} and~\eqref{eq_def_corr}, respectively. Similarly, if we were to start our system with a collection of $N$ neutrons, each having a source distribution ${\cal Q}_n(x_0, t_0)$, and $M$ precursors, each having a source distribution ${\cal Q}_m(x_0, t_0)$, then Eq.~\eqref{pal_bell_W_source} would be replaced by
\begin{equation}
W_{{\cal Q}_{n+m}}(z_1,z_2) = \left[ \int^{t_1}_{-\infty} dt_0\int dx_0 \, W(z_1,z_2,x_0,t_0) {\cal Q}_n(x_0, t_0) \right]^{N} \left[ \int^{t_1}_{-\infty} dt_0\int dx_0 \, W(z_1,z_2,x_0,t_0) {\cal Q}_m(x_0, t_0) \right]^{M}.
\label{pal_bell_W_source_NM}
\end{equation}
Correspondingly, setting ${\cal N} = N + M$ and denoting
\begin{equation}
\mathcal{Q}_{n+m}(x_0, t_0) = \frac{N}{N+M} \mathcal{Q}_{n}(x_0, t_0) + \frac{M}{N+M} \mathcal{Q}_{m}(x_0, t_0),
\end{equation}
for the average number of particles we would have
\begin{equation}
\mathbb{E}[n_1](t_1|{\cal Q}_{n+m}) = {\cal N} \int^{t_1}_{-\infty}dt_0 \int d x_0 \,\mathbb{E}[n_1](t_1 | x_0, t_0)  {\cal Q}_{n+m}(x_0,t_0) ,
\label{eq:pal_bell_ave_appendix_solution_source_NM}
\end{equation}
and for the pair correlation we would have
\begin{align}
&\mathbb{E}[n_1 n_2](t_1,t_2 |{\cal Q}_{n+m}) = {\cal N}^2 \mathbb{E}[n_1](t_1|{\cal Q}_{n+m}) \mathbb{E}[n_2](t_2|{\cal Q}_{n+m}) \nonumber\\
&\qquad - N \mathbb{E}[n_1](t_1|{\cal Q}_{n}) \mathbb{E}[n_2](t_2|{\cal Q}_{n}) -M \mathbb{E}[n_1](t_1|{\cal Q}_{m}) \mathbb{E}[n_2](t_2|{\cal Q}_{m})\nonumber\\
&\qquad+ {\cal N} \int^{t_1}_{-\infty}dt_0 \int d x_0 \mathbb{E}[n_1 n_2](t_1,t_2 | x_0, t_0)  {\cal Q}_{n+m}(x_0,t_0) .
\label{eq:pal_bell_cov_appendix_solution_source_NM}
\end{align}
The P\'al and Bell backward formalism can be easily extended to include more general detectors (for instance, an array $\textbf{V} = (V_1,\dots,V_K)$ of $K$ non-overlapping regions), and to more general sources (for instance, $N$ and $M$ can be themselves random variables with Poisson distributions).

\section{Solutions for the anarchic model}
\label{appendix:anarch_sol}

For the critical anarchic model with a critical source, the solution of the
pair correlation function in Eq.~\eqref{eq:pal_bell_cov_appendix} can be
derived explicitly and is provided in this section for reference. For
$t_1<t_2$, the pair correlation function $u(x_1,t_1,x_2,t_2)$ can be decomposed
as follows:
\begin{align}
    u(x_1,t_1,x_2,t_2) &= \frac{{\cal N}({\cal N}-1)\theta^2}{4 L^2 (1+\theta)^2} + \frac{{\cal N} \theta}{2 L(1+\theta)}{\cal G}(x_2,t_2,x_1,t_1) \nonumber \\
    &\qquad + \frac{{\cal N} \theta}{2L(1+\theta)} \left[ u_{pp}(x_1,t_1,x_2,t_2) + u_{pd}(x_1,t_1,x_2,t_2) + u_{dp}(x_1,t_1,x_2,t_2) + u_{dd}(x_1,t_1,x_2,t_2) \right].
    \label{eq:first_anarch_analytical}
\end{align}
Each component is defined in terms of the respective eigenmode expansion, namely,
\begin{align}
    u_{pp} &=  \sum_{k} \varphi_k(x_1) \varphi_k^\dagger(x_2) u_{pp}^k \\
    u_{pd} &=  \sum_{k} \varphi_k(x_1) \varphi_k^\dagger(x_2) u_{pd}^k \\
    u_{dp} &=  \sum_{k} \varphi_k(x_1) \varphi_k^\dagger(x_2) u_{dp}^k \\
    u_{dd} &=  \sum_{k} \varphi_k(x_1) \varphi_k^\dagger(x_2) u^k_{dd}.
\end{align}
For the prompt-prompt component $u_{pp}^k$ we have:
\begin{align}
    u_{pp}^0(t_1,t_2) = \frac{\beta \nu_{p,2}}{(1+\theta)^2}\left[\theta^2t_1 + \theta\left(\frac{e^{\omega_d t_1}-1}{\omega_d} + \frac{e^{\omega_d t_2} - e^{\omega_d(t_2-t_1)}}{\omega_d}\right) + \frac{e^{\omega_d(t_1+t_2)}-e^{\omega_d(t_2-t_1)}}{2\omega_d}\right],
    \label{eq:upp_0}
\end{align}
where we have introduced $\omega_d = -\tau_2^{-1}= - \beta \nu_{d,1} - \lambda$, and
\begin{align}
    u_{pp}^k(t_1,t_2) &= \frac{\beta \nu_{p,2}}{(\omega_k^{+} -\omega_k^{-})^2}\left[\frac{(\omega_k^{+} + \lambda)^2}{2\omega_k^{+}}\left(e^{\omega_k^{+}(t_1+t_2)} - e^{\omega_k^{+}(t_2-t_1)}\right) + \frac{(\omega_k^{-}+\lambda)^2}{2\omega_k^{-}}\left(e^{\omega_k^{-}(t_1+t_2)} - e^{\omega_k^{-}(t_2-t_1)}\right)\vphantom{\frac{\varphi_k(x_1)}{(\omega_k^{+})^2}} \nonumber\right.\\
           & \qquad \left. + \frac{(\omega_k^{+}+\lambda)(\omega_k^{-}+\lambda)}{\omega_k^{+}+\omega_k^{-}}\left(e^{\omega_k^{+}(t_2-t_1)} + e^{\omega_k^{-}(t_2-t_1)} - (e^{\omega_k^{+}t_1 + \omega_k^{-}t_2} + e^{\omega_k^{+}t_2 + \omega_k^{-}t_1})\right) \nonumber \right],
           \label{eq:upp}
\end{align}
for $k \ge 1$. For the prompt-delayed component $u_{pd}^k$ we have
\begin{align}
    u_{pd}^0(t_1,t_2) &= \frac{\beta \nu_{d,1} \nu_{p,1} \theta}{(1+\theta)^2}\left[\theta t_1 + \frac{e^{\omega_d t_1} - 1}{\omega_d} + \frac{\theta(e^{\omega_d(t_2-t_1)} - e^{\omega_d t_2})}{\omega_d} + \frac{e^{\omega_d(t_2-t_1)} - e^{\omega_d(t_1+t_2)}}{2\omega_d}\right]
\end{align}
and
\begin{align}
    u_{pd}^k(t_1,t_2) &=  \frac{\beta \lambda \nu_{d,1}\nu_{p,1}}{(\omega_k^{+} -\omega_k^{-})^2}\left[(\omega_k^{+}+\lambda)\left(\frac{e^{\omega_k^{+}(t_1+t_2)}-e^{\omega_k^{+}(t_2-t_1)}}{2\omega_k^{+}} \right. + \frac{e^{\omega_k^{-}(t_2-t_1)}-e^{\omega_k^{+}t_1 + \omega_k^{-}t_2}}{\omega_k^{+} + \omega_k^{-}}\right) \nonumber \\
           &\qquad + (\omega_k^{-}+\lambda)\left(\frac{e^{\omega_k^{+}(t_2-t_1)} - e^{\omega_k^{+}t_2 + \omega_k^{-}t_1}}{\omega_k^{+} + \omega_k^{-}}  + \frac{e^{\omega_k^{-}(t_1+t_2)} - e^{\omega_k^{-}(t_2-t_1)}}{2\omega_k^{-}}\left.\vphantom{\frac{\varphi_k(x_1)}{(\omega_k^{+})^2}}\right)\right]
\end{align}
for $k \ge 1$. For the delayed-prompt component $u_{dp}^k$ we have
\begin{align}
    u_{dp}^0(t_1,t_2) = \frac{\beta\nu_{d,1}\nu_{p,1}\theta}{(1+\theta)^2}\left[\theta t_1 + \theta\left(\frac{1-e^{\omega_d t_1}}{\omega_d}\right) + \frac{e^{\omega_d t_2} - e^{\omega_d(t_2-t_1)}}{\omega_d} + \frac{e^{\omega_d(t_1-t_2)}-e^{\omega_d(t_2+t_1)}}{2\omega_d}\right]
\end{align}
and
\begin{align}
    u_{dp}^k(t_1,t_2) &= \frac{\beta\lambda\nu_{d,1}\nu_{p,1}}{(\omega_k^{+} -\omega_k^{-})^2}\left[(\omega_k^{+}+\lambda)\left(\frac{e^{\omega_k^{+}(t_1+t_2)}-e^{\omega_k^{+}(t_2-t_1)}}{2\omega_k^{+}} \nonumber\right. + \frac{e^{\omega_k^{+}(t_2-t_1)}-e^{\omega_k^{+}t_2 + \omega_k^{-}t_1}}{\omega_k^{+} + \omega_k^{-}}\right)\left.\vphantom{\frac{\varphi_k(x_1)}{(\omega_k^{+})^2}}\right. \nonumber \\
           &\qquad + (\omega_k^{-}+\lambda)\left(\frac{e^{\omega_k^{-}(t_2-t_1)} - e^{\omega_k^{+}t_1 + \omega_k^{-}t_2}}{\omega_k^{+} + \omega_k^{-}} + \frac{e^{\omega_k^{-}(t_1+t_2)} - e^{\omega_k^{-}(t_2-t_1)}}{2\omega_k^{-}}\left.\vphantom{\frac{\varphi_k(x_1)}{(\omega_k^{+})^2}}\right)\right]
\end{align}
for $k \ge 1$. Finally, for the delayed-delayed component $u_{dd}^k$ we have
\begin{align}
    u^0_{dd}(t_1,t_2) &= \frac{\beta \nu_{d,2}\theta^2}{(1+\theta)^2}\left[t_1 + \frac{1+e^{\omega_d( t_1 - t_2)} - e^{\omega_d t_1} - e^{\omega_d t_2}}{\omega_d}
    + \frac{e^{\omega_d(t_1+t_2)} - e^{\omega_d(t_2-t_1)}}{2\omega_d}\vphantom{\frac{\left(e^{\omega_d(t_2-t_1)} - 1\right)\left(1-e^{-\lambda t_1}\right)}{\lambda}}\right]
    \label{eq:last_anarch_analytical}
\end{align}
and
\begin{align}
    u_{dd}^k(t_1,t_2) &=  \frac{\beta\lambda^2\nu_{d,2}}{(\omega_k^{+} - \omega_k^{-})^2} \left[\frac{e^{\omega_k^{+}(t_1+t_2)} - e^{\omega_k^{+}(t_2-t_1)}}{2 \omega_k^{+}} + \frac{e^{\omega_k^{-}(t_1+t_2)} - e^{\omega_k^{-}(t_2-t_1)}}{2\omega_k^{-}} \right. \nonumber\\
           &\qquad \left. + \frac{e^{\omega_k^{+}(t_2-t_1)} + e^{\omega_k^{-}(t_2-t_1)} - e^{\omega_k^{+}t_1 + \omega_k^{-}t_2} - e^{\omega_k^{-}t_1 + \omega_k^{+}t_2}}{\omega_k^{+}+\omega_k^{-}} \right]
\end{align}
for $k \ge 1$.  Note that Eq.~\eqref{eq_def_corr} of the main text, which is
applicable to generic values of $t_1$ and $t_2$, is recovered by operating the
substitutions $t_1\to \min\{t_1,t_2\}$ and $t_2\to\max\{t_1,t_2\}$ in
Eq.~\eqref{eq:first_anarch_analytical}.

Taking $t_1 = t_2 = t$, we can obtain an expression for the mean-squared pair distance $\langle r^2 \rangle(t)$. For our case (reflection boundary conditions,  $\nu_{d,2} = 0$), the expression reads:
\begin{align}
    \langle r^2 \rangle(t) &= \frac{\displaystyle\iint (x-y)^2 u(x,y,t)\,dx\,
    dy}{u(t,t)}\nonumber\\
    &= \frac{2 L^2}{3} \left( 1 - \frac{1}{ N + u_0^{pp}(t) + u_0^{dp}(t) + u_{pd}^0(t)}\right)
    - \frac{64 L^2}{ N + u_{pp}^0(t) + u_{dp}^0(t) + u_{pd}^0(t) }  \sum_{\substack{k=1\\k \text{ odd}}}^{+\infty} \frac{u^k_{pp}(t) + u^k_{dp}(t) + u_{pd}^k(t)}{(k \pi)^4}
    \label{eq:anarch_pair_distance}
\end{align}

\section{Forward formalism for the models with population control}
\label{appendix:master_equations}

When population control is enforced, particle histories are no longer independent
and we have to resort to the forward formalism.
Using the strategy proposed in
\cite{houchmandzadehNeutronFluctuationsImportance2015}, we partition the viable
reactor space into a set of $K$ equal segments of length $\delta x = 2L/K$,
which we denote as $V_k$, with $k\in\{1,\ldots, K\}$. Correspondingly, we
write $\textbf{n} = (n_1,\dots,n_K)$, $\textbf{m} = (m_1,\dots,m_K)$, and we
denote by ${\cal P}(\textbf{n},\textbf{m} , t)$ the
joint probability of observing $n_k$ neutrons and $m_k$ precursors in each of
the $K$ detectors at time $t$, for $1\leq k\leq K$,
given an initial condition with $N$ neutrons and $M$
precursors at time
$t_0=0$ uniformly distributed in space. The stochastic rules and
the associated transition rates are those defined in
Sec.~\ref{sec:population_control}. The evolution equations for ${\cal
P}(\textbf{n},\textbf{m}, t)$ can be written more concisely introducing the
annihilation $a_k$ and creation $a_k^{\dagger}$ operators, whose action on a
state vector $\textbf{v} = (v_1,\dots,v_K)$ is defined by
\begin{align}
    a_k \textbf{v} &= (\dots,v_{k-1},v_k - 1,v_{k+1}, \dots) \\
    a_k^{\dagger}\textbf{v} &= (\dots, v_{k-1},v_k + 1,v_{k+1}, \dots).
\end{align}

\subsection{Master equations}

For the \emph{$N$-control} model described in Sec.~\ref{sss:N_cst}, where exactly two neutrons and at most one precursor are produced in each fission event, and population control is enforced on neutrons alone, probability balance yields the master equation
\begin{align}
    \frac{\partial}{\partial t} {\cal P}(\textbf{n}, \textbf{m}, t) &= \sum_{i} \left[\vphantom{\sum_j}\xi(n_{i+1}+1){\cal P}(a_{i+1}^{\dagger} a_i\textbf{n}, \textbf{m}, t) + \xi(n_{i-1}+1){\cal P}(a_{i-1}^{\dagger} a_i\textbf{n}, \textbf{m} ,t) - 2\xi n_{i}{\cal P}(\textbf{n}, \textbf{m} , t) \right] \nonumber \\
    & + \beta (1- \nu_{d,1})N \sum_{i,j \neq i}\left( \frac{n_i+1}{N}\frac{n_j-1}{N-1} {\cal P}(a_i^{\dagger} a_{j}\textbf{n}, \textbf{m} , t)- \frac{n_i}{N}\frac{n_j}{N-1} {\cal P}(\textbf{n}, \textbf{m},t)\right) \nonumber\\
    & + \beta  \nu_{d,1} N \sum_{i,j \neq i} \left( \frac{n_i+1}{N}\frac{n_j-1}{N-1}{\cal P}(a_i^{\dagger} a_j \textbf{n}, a_j\textbf{m},t) - \frac{n_i}{N}\frac{n_j}{N-1}{\cal P}(\textbf{n}, \textbf{m},t)\right) \nonumber \\
    & + \beta  \nu_{d,1} N\sum_{i} \left( \frac{n_i}{N}\frac{n_i-1}{N-1}{\cal P}(\textbf{n}, a_j\textbf{m},t) -  \frac{n_i}{N}\frac{n_i-1}{N-1}{\cal P}(\textbf{n}, \textbf{m},t)\right) \nonumber \\
    & + \lambda m\sum_{i,j\neq i} \left( \frac{m_i+1}{m}\frac{n_j+1}{N} {\cal P}(a_i a_j^{\dagger}\textbf{n}, a_i^{\dagger}\textbf{m},t) - \frac{m_i}{m}\frac{n_j}{N} {\cal P}(\textbf{n}, \textbf{m},t)\right) \nonumber \\
    & + \lambda m\sum_{i} \left( \frac{m_i+1}{m}\frac{n_i}{N} {\cal P}(\textbf{n}, a_i^{\dagger}\textbf{m},t) -  \frac{m_i}{m}\frac{n_i}{N} {\cal P}(\textbf{n}, \textbf{m},t)\right).
    \label{eq:master_partial}
\end{align}
where $\xi$ is the diffusion rate of a neutron from one site to a neighbouring one, $m = \sum_i m_i$ is the total number of precursors, and the total number of neutrons is $n = \sum_i n_i = N$ because of population control. As a special case, when precursors are neglected ($\lambda=0$, $\nu_{d,1}=0$),
Eq.~\eqref{eq:master_partial} degenerates into
\begin{align}
    \frac{\partial}{\partial t} {\cal P}(\textbf{n} , t) &= \sum_{i} \left[\vphantom{\sum_j}\xi(n_{i+1}+1){\cal P}(a_{i+1}^{\dagger} a_i\textbf{n},t) + \xi(n_{i-1}+1){\cal P}(a_{i-1}^{\dagger} a_i\textbf{n},t) - 2\xi n_{i}{\cal P}(\textbf{n},t) \right.\nonumber \\
    & + \left.\sum_{j\neq i}\left(\beta N \frac{n_i+1}{N}\frac{n_j-1}{N-1} {\cal P}(a_i^{\dagger} a_{j}\textbf{n},t)-\beta N \frac{n_i}{N}\frac{n_j}{N-1} {\cal P}(\textbf{n},t)\right) \nonumber \right],
    \label{eq:master_prompt}
\end{align}
which corresponds to the population control model proposed in
Ref.~\citenum{meyerClusteringIndependentlyDiffusing1996} and revisited in
Ref.~\citenum{mulatierCriticalCatastropheRevisited2015}. From
Eq.~\eqref{eq:master_partial} the master equation for the total population sizes is easily derived, and, because of strict population control on $N$, it is an equation on $m$ only, namely
\begin{equation}
    \frac{\partial}{\partial t}{\cal P}(m,t) = \beta N  \nu_{d,1} {\cal P}(m-1,t) - \beta N  \nu_{d,1} {\cal P}(m,t) + (m+1)\lambda {\cal P}(m+1,t) - m \lambda {\cal P}(m,t).
    \label{eq:master_integrated_partial}
\end{equation}
By taking the moments of Eq.~\eqref{eq:master_integrated_partial}, one derives
Eqs.~\eqref{eq:eqdiff_mtot_cons_n} and Eqs.~\eqref{eq:eqdiff_uvwtot_cons_n} of the main
text.

For the \emph{$NM$-control} model described in Sec.~\ref{sss:NM_cst}, where population control is enforced on neutrons and precursors, probability balance yields the master equation
\begin{align}
    \frac{\partial}{\partial t} {\cal P}(\textbf{n}, \textbf{m},t) &= \sum_{i} \left[\vphantom{\sum_j}\xi(n_{i+1}+1){\cal P}(a_{i+1}^{\dagger} a_i\textbf{n}, \textbf{m},t) + \xi(n_{i-1}+1){\cal P}(a_{i-1}^{\dagger} a_i\textbf{n}, \textbf{m},t) - 2\xi n_{i}{\cal P}(\textbf{n}, \textbf{m},t) \right.\nonumber \\
    & + \left.\sum_{j \neq i}\left(\frac{\beta (1- \nu_{d,1})}{N-1} (n_i+1)(n_j-1) {\cal P}(a_i^{\dagger} a_{j}\textbf{n}, \textbf{m},t)-\frac{\beta (1- \nu_{d,1})}{N-1} n_i n_j {\cal P}(\textbf{n}, \textbf{m},t)\right)\right] \nonumber\\
     + \frac{\beta  \nu_{d,1}}{(N-1)M}&\left[\sum_{\substack{i,j,k \\ i\neq j \neq k}} \left( (n_i+1)(n_j-1)(m_k+1){\cal P}(a_i^{\dagger} a_j \textbf{n}, a_j a_k^{\dagger}\textbf{m},t) -  n_i n_j m_k{\cal P}(\textbf{n}, \textbf{m},t)\right)\right. \nonumber \\
    & \qquad\sum_{\substack{i,k \\ i \neq k}} \left( n_i(n_i-1)(m_k+1){\cal P}(\textbf{n}, a_i a_k^{\dagger}\textbf{m},t) -  n_i (n_i-1) m_k{\cal P}(\textbf{n}, \textbf{m},t)\right) \nonumber \\
    & \qquad\sum_{\substack{i,j \\ i \neq j}} \left( (n_i+1)(n_j-1)(m_i+1){\cal P}(a_i^{\dagger} a_j \textbf{n}, a_i^{\dagger} a_j\textbf{m},t) -  n_i n_j m_i{\cal P}(\textbf{n}, \textbf{m},t)\right) \nonumber \\
    & \qquad\left.\sum_{\substack{i,j \\ i \neq j}} \left( (n_i+1)(n_j-1)m_j{\cal P}(a_i^{\dagger} a_j \textbf{n},\textbf{m},t) -  n_i n_j m_j{\cal P}(\textbf{n}, \textbf{m},t)\right)\right] \nonumber \\
     + \frac{\lambda}{N}&\sum_{\substack{i,j \\ i \neq j}} \left(m_i (n_j+1){\cal P}(a_i a_j^{\dagger}\textbf{n}, \textbf{m},t) - m_i n_j {\cal P}(\textbf{n}, \textbf{m},t)\right).
     \label{eq:master_full}
\end{align}
with $n = N$ and $m = M$ at all times. Finally, for the {\em immigration model} described in Sec.~\ref{sss:imm}, probability balance yields the master equation
\begin{multline}
    \frac{\partial}{\partial t} {\cal P}(\textbf{n}, t) = \sum_{i} \left[\vphantom{\sum_j}\xi(n_{i+1}+1){\cal P}(a_{i+1}^{\dagger} a_i\textbf{n},t) + \xi(n_{i-1}+1){\cal P}(a_{i-1}^{\dagger} a_i\textbf{n},t) - 2\xi n_{i}{\cal P}(\textbf{n},t) \right. \nonumber \\
    + \frac{\beta}{N-1}\sum_{j\neq i}\left((n_i+1)(n_j-1) {\cal P}(a_i^{\dagger} a_{j}\textbf{n},t)- n_i n_j {\cal P}(\textbf{n},t)\right)
    \left. + \frac{\lambda M}{N} \sum_{j\neq i} \mathcal{Q}_j\left( (n_i+1)\mathcal{P}(a_i^{\dagger} a_{j}\textbf{n},t) -n_i\mathcal{P}(\textbf{n},t)\right) \right]
    \text.
    \label{eq:master_immigration}
\end{multline}

\subsection{Moment equations}

Once the master equations for the probability ${\cal P}(\textbf{n},\textbf{m} , t)$ have been established, the moments of the population are derived by summation: for the average densities we have
\begin{subequations}
\begin{align}
    \begin{split}
        \mathbb{E}[n_i](t)=
\sum\limits_{\textbf{n}, \textbf{m}} n_i\,\mathcal{P}(\textbf{n}, \textbf{m},t)
    \end{split} \\
    \begin{split}
        \mathbb{E}[m_i](t)=
\sum\limits_{\textbf{n}, \textbf{m}} m_i\,\mathcal{P}(\textbf{n}, \textbf{m},t)
\text,
    \end{split}
\end{align}%
\label{eq:average_pop_nm}%
\end{subequations}%
for $i \in \{ 1, \ldots, K \}$. The two-point correlations between particles detected in $V_i$ and particles detected in $V_j$ at time $t$ are similarly obtained from
\begin{subequations}
    \begin{align}
        \begin{split}
            \mathbb{E}[n_i n_j](t)=
            \sum\limits_{\textbf{n}, \textbf{m}} n_i\,n_j\,\mathcal{P}(\textbf{n}, \textbf{m}, t)
        \end{split}\\
        \begin{split}
            \mathbb{E}[n_i m_j](t)=
            \sum\limits_{\textbf{n},\textbf{m}} n_i\,m_j\,\mathcal{P}(\textbf{n},\textbf{m},t)
        \end{split}\\
        \begin{split}
            \mathbb{E}[m_i m_j](t)=
            \sum\limits_{\textbf{n},\textbf{m}} m_i\,m_j\,\mathcal{P}(\textbf{n},\textbf{m},t) \text,
        \end{split}
    \end{align}%
    \label{eq:correlations_pop_nm}%
\end{subequations}%
for $i,j \in \{ 1, \ldots, K \}$. For the \emph{$N$-control} model, the discrete moment equations obtained directly from Eq.~\eqref{eq:master_partial} read
\begin{subequations}
    \begin{align}
        \begin{split}
            \frac{\partial}{\partial t} \mathbb{E}[n_i](t) = \xi \Delta \mathbb{E}[n_i](t) + \lambda \left(\mathbb{E}[m_i](t) - \frac{\mathbb{E}[n_i m](t)}{N} \right)
        \end{split}\\
        \begin{split}
            \frac{\partial}{\partial t}\mathbb{E}[m_i](t) = \beta \nu_{d,1} \mathbb{E}[n_i](t) - \lambda \mathbb{E}[m_i](t)
        \end{split}
    \end{align}
\end{subequations}
for the first spatial moment, and 
\begin{align}
    \frac{\partial}{\partial t} \mathbb{E}[n_i n_j](t) &= \xi \left(\mathbb{E}[n_i \Delta n_{j}](t) + \mathbb{E}[n_j\Delta n_{i}](t) \right) - \frac{2 \beta}{N_0 -1}\mathbb{E}[n_i n_j](t) + \lambda C_N \left(\mathbb{E}[n_i m_j](t) + \mathbb{E}[m_i n_j](t)\right) \nonumber \\ 
    & - \frac{2\lambda}{N}\mathbb{E}[n_i n_j m](t)
    +\frac{2 \beta}{N_0 -1}\delta_{i,j}N_0\mathbb{E}[n_i](t) + \lambda\delta_{i,j}\left( \mathbb{E}[m_i](t) + \frac{\mathbb{E}[n_i m](t)}{N}\right) \nonumber \\
    & + \delta_{i,j}\xi\left(\mathbb{E}[n_{i+1}](t) + \mathbb{E}[n_{i-1}](t) + 2\mathbb{E}[n_i](t) \right)
    - \delta_{i+1,j} \xi\left(\mathbb{E}[n_{i+1}](t) + \mathbb{E}[n_i](t) \right) \nonumber\\
    & - \delta_{i-1,j}\xi\left(\mathbb{E}[n_i](t) + \mathbb{E}[n_{i-1}](t) \right) \\
    \frac{\partial}{\partial t} \mathbb{E}[n_i m_j](t) &= \xi \mathbb{E}[m_j \Delta n_i ](t) - \lambda \frac{N-1}{N} \mathbb{E}[n_i m_j](t) - \lambda \frac{\mathbb{E}[n_i m_j m](t)}{N} \nonumber \\ 
    & + \beta \nu_{d,1} C_{N-1} \mathbb{E}[n_i n_j](t) + \lambda \mathbb{E}[m_i m_j](t) + \delta_{i,j}\left( \frac{\beta \nu_{d,1} N}{N-1}\mathbb{E}[n_i](t) - \lambda \mathbb{E}[m_i](t) \right) \\
    \frac{\partial}{\partial t} \mathbb{E}[m_i m_j](t) &= \beta \nu_{d,1} \left(\mathbb{E}[m_i n_j](t) + \mathbb{E}[n_i m_j](t) \right) - 2\lambda\mathbb{E}[m_i m_j](t) + \delta_{i,j}\left(\beta \nu_{d,1} \mathbb{E}[n_i](t) + \lambda \mathbb{E}[m_i](t) \right)
    \label{eq:conservation_n_second_moment_discretized}
\end{align}
for the second spatial moment, where we define $\Delta f_i = f_{i+1} -2f_i + f_{i+1}$. 

For the \emph{$NM$-control model}, we similarly obtain equations for the first spatial
moments, namely
\begin{align}
    \frac{\partial}{\partial t} \mathbb{E}[n_i](t) &= \xi \Delta \mathbb{E}[n_i](t) + \lambda \left(\mathbb{E}[m_i](t) - \mathbb{E}[n_i](t) \frac{M}{N} \right) \\
    \frac{\partial}{\partial t} \mathbb{E}[m_i](t) &= \beta \nu_{d,1} \left(\mathbb{E}[n_i](t) - \mathbb{E}[m_i](t) \frac{N}{M} \right)\text.
\end{align}
The equations for the second spatial moment read
\begin{align}
    \frac{\partial}{\partial t} \mathbb{E}[n_i n_j](t) &= \xi \left(\mathbb{E}[n_i \Delta n_j](t) + \mathbb{E}[n_j \Delta n_i](t) \right) - \left(\frac{2 \beta }{N -1} + \frac{2 \lambda M}{N}\right)\mathbb{E}[n_i n_j](t) + \lambda C_N(\mathbb{E}[m_i n_j](t) + \mathbb{E}[n_i m_j](t)) \nonumber\\
    & + \frac{2 N \beta}{N -1}\delta_{i,j}\mathbb{E}[n_i](t) + \lambda \delta_{i,j}\left(\mathbb{E}[m_i](t) + \mathbb{E}[n_i](t) \frac{M}{N} \right) \nonumber \\
    & + \delta_{i,j}\xi\left(\mathbb{E}[n_{i+1}](t) + \mathbb{E}[n_{i-1}](t) + 2\mathbb{E}[n_i](t) \right)
    - \delta_{i+1,j} \xi\left(\mathbb{E}[n_{i+1}](t) + \mathbb{E}[n_i](t) \right) \nonumber \\
    &- \delta_{i-1,j}\xi\left(\mathbb{E}[n_i](t) + \mathbb{E}[n_{i-1}](t) \right)
    \label{eq:conservation_n_m_second_moment_discretized}\\
    \frac{\partial}{\partial t} \mathbb{E}[n_i m_j](t) &= \xi \mathbb{E}[ m_j \Delta n_i](t) - \left(\frac{\beta \nu_{d,1} N}{M} + \lambda \frac{M}{N}\right) \mathbb{E}[n_i m_j](t) + \beta \nu_{d,1} \frac{ N-2}{N-1} \mathbb{E}[n_i n_j](t) + \lambda \mathbb{E}[m_i m_j](t) \nonumber \\
    & \qquad+ \frac{\beta \nu_{d,1} N}{(N-1)} \delta_{i,j}\mathbb{E}[n_i](t) \\
    \frac{\partial}{\partial t} \mathbb{E}[m_i m_j](t) &= \beta \nu_{d,1} C_M\left(\mathbb{E}[m_i n_j](t) + \mathbb{E}[n_i m_j](t) \right) - \frac{2 \beta \nu_{d,1} N}{M} \mathbb{E}[m_i m_j](t) \nonumber + \beta \nu_{d,1}\delta_{i,j} \left( \mathbb{E}[n_i](t) + \mathbb{E}[m_i](t) \frac{N}{M} \right)\text.
\end{align}

Finally, for the immigration model, the first spatial moment satisfies
\begin{equation}
    \frac{\partial}{\partial t} \mathbb{E}[n_i](t) = \xi  \Delta \mathbb{E}[n_i](t)
    - \frac{\lambda M}{N}\mathbb{E}[n_i](t)+\lambda M \mathcal{Q}_i\text,
    \label{eq:immigration_first_moment_discretized}
\end{equation}
where $\mathcal{Q}_i$ is the intensity of the source in the $i$-th detector
region. The equation for the second spatial moment reads
\begin{align}
    \frac{\partial}{\partial t} \mathbb{E}[n_i n_j](t) &= \xi \left(\mathbb{E}[n_i \Delta n_j](t) + \mathbb{E}[\Delta n_i n_j](t) \right) -\left(\frac{2 \beta}{N-1} + \frac{2\lambda M}{N}\right) \mathbb{E}[n_i n_j](t) + \delta_{i,j}\frac{2 N \beta}{N-1} \mathbb{E}[n_i](t) \nonumber \\
    &+ \lambda M\delta_{i,j}
    \left(\mathcal{Q}_i+\frac{\mathbb{E}[n_i](t)}{N}\right)
    + \lambda M \frac{N-1}{N} \left(\mathcal{Q}_j\mathbb{E}[n_i](t) + \mathcal{Q}_i\mathbb{E}[n_j](t) \right)\nonumber\\
    & + \delta_{i,j}\xi\left(\mathbb{E}[n_{i+1}](t) + \mathbb{E}[n_{i-1}](t) + 2\mathbb{E}[n_{i}](t) \right)
    - \delta_{i+1,j} \xi\left(\mathbb{E}[n_{i+1}](t) + \mathbb{E}[n_{i}](t) \right) \nonumber\\
    &- \delta_{i-1,j}\xi\left(\mathbb{E}[n_{i}](t) + \mathbb{E}[n_{i-1}](t) \right)
    \text.
    \label{eq:immigration_second_moment_discretized}
\end{align}

Often, it is preferable to work with continuous observables: we therefore define the neutron and precursor densities by taking the limit
\begin{align}
     n(x,t|x_0,t_0) &=\lim_{\delta x \to 0} \frac{\mathbb{E}[n_i](t|x_0,t_0)}{\delta x} \\
     m(x,t|x_0,t_0) &=\lim_{\delta x \to 0} \frac{\mathbb{E}[n_i](t|x_0,t_0)}{\delta x},
\end{align}
Similarly, we define the pair correlation
functions by taking the limits
\begin{subequations}
    \begin{align}
        \begin{split}
            u(x,y,t|x_0,t_0) = \lim_{\delta x \to 0} \frac{\mathbb{E}[n_i n_j](t|x_0,t_0)  }{\delta x ^ 2}
        \end{split}\\
        \begin{split}
            v(x,y,t|x_0,t_0) = \lim_{\delta x \to 0} \frac{\mathbb{E}[n_i m_j](t|x_0,t_0)  }{\delta x ^ 2}
        \end{split}\\
        \begin{split}
            w(x,y,t|x_0,t_0) = \lim_{\delta x \to 0} \frac{\mathbb{E}[m_i m_j](t|x_0,t_0)  }{\delta x ^ 2},
        \end{split}%
    \end{align}%
\end{subequations}%
which describe the correlations at time $t$ between a particle found at $x$ and a particle found at $y$.
The $i$-th ($j$-th) detector is defined to be the one containing $x$ ($y$).
We also define the diffusion coefficient to be
\[
{\cal D} = \lim_{\delta x \to 0} (\xi\, \delta x^2)
\text.
\]

\section{\texorpdfstring{Solutions of the $NM$-control model}{Solutions of the NM-control model}}
\label{appendix:analytic_solutions}

We can obtain solutions to Eqs.~\eqref{eq:NM_cst_2nd} by using a Fourier decomposition which, for reflection boundary conditions, reduces to
\begin{equation*}
    f_{\infty} = \sum_{k=-\infty}^{+\infty} f_{k,k} \cos\left(\frac{k\pi}{2L}(L-x)\right) \cos\left(\frac{k\pi}{2L}(L-y)\right).
\end{equation*}
Indeed, in this case all the coefficients with $k_1 \neq k_2$ are zero. The terms $f_{k,k}$ thus read
\begin{align}
    \Tilde{u}_{k,k} &= \frac{m(N-1)\lambda^2 + n\beta\left(N(\alpha_1+\Tilde{k}^2 D) + 2(N-1) \nu_{d,1} \lambda\right)}{L N\left(\alpha_1 + \Tilde{k}^2 D\right)\left(\tau_{n} + 2 \Tilde{k}^2 D\right)-2L(N-2)\beta \nu_{d,1}\lambda}\\
    v_{k,k} &= \frac{2 n N \beta \left((N-1)\tau_{n} + (N-2)\beta + 2\Tilde{k}^2(N-1) D\right) \nu_{d,1} + m(N-1)N\left(\tau_{n} + 2\Tilde{k}^2 D\right)\lambda}{2 L(N-1)\left(N(\alpha_1 + \Tilde{k}^2 D)(\tau_{n}+2\Tilde{k}^2 D\right)-2(N-2)\beta \nu_{d,1}\lambda)} \\
    \Tilde{w}_{k,k} &= \frac{M-1}{N}\Tilde{v}_{k,-k}
    \label{eq:NM_sol}
\end{align}
where $\tau_{n}$, $\tau_c$ are given by Eqs.~\eqref{eq:nm_constants}, and $n$, $m$ and $\tilde{k}$ are respectively given by
\begin{equation*}
    n = \frac{N}{2 L} \text, \qquad m = \frac{M}{2 L}, \qquad \Tilde{k} = \frac{k \pi}{2 L}.
\end{equation*}

\end{widetext}

\bibliography{betterbiblio}

\end{document}